\documentclass[a4paper,fleqn,usenatbib]{mnras}

\usepackage{mathptmx}
\usepackage[T1]{fontenc}
\usepackage{ae,aecompl}
\usepackage{graphicx}	% Including figure files
\usepackage{amsmath}	% Advanced maths commands
\usepackage{amssymb}	% Extra maths symbols

\title[Chemical abundances in the GC NGC6723]{Chemical abundances in the metal-intermediate GC NGC6723}
\author[J. Crestani et al.]{
Juliana Crestani,$^{1,2,3}$\thanks{E-mail: juliana.crestani@ufrgs.br}
Alan Alves-Brito,$^{1}$
Giuseppe Bono,$^{2,3}$
Arthur A. Puls,$^{1,4}$
\newauthor
\hspace{0.02 cm} and Javier Alonso-Garc\'ia$^{5,6}$
\\
$^{1}$Departamento de Astronomia, Universidade Federal do Rio Grande do Sul, Av. Bento Gon\c{c}alves 6500, Porto Alegre 91501-970, Brazil \\
$^{2}$Universit\`a di Roma Tor Vergata, Via della Ricerca Scientifica 1, I-00133 Roma, Italy \\
$^{3}$INAF--Osservatorio Astronomico di Roma, Via Frascati 33, I-00040 Monteporzio Catone, Italy \\
$^{4}$Research School of Astronomy and Astrophysics, Australian National University, Canberra, ACT 2611, Australia \\
$^{5}$Centro de Astronom\'ia (CITEVA), Universidad de Antofagasta, Av. Angamos 601, Antofagasta, Chile \\
$^{6}$Millennium Institute of Astrophysics, Santiago, Chile \\
}

\date{Accepted XXX. Received YYY; in original form ZZZ}
\pubyear{2019}

\begin{document}
\label{firstpage}
\pagerange{\pageref{firstpage}--\pageref{lastpage}}
\maketitle

\begin{abstract}
We have performed a detailed spectral analysis of the inner halo Galactic globular cluster (GC) NGC~6723 using high resolution (R$\approx$ 22000--48000) spectra for for eleven red giant branch stars collected with MIKE (Magellan) and FEROS (MPG/ESO). This globular is located at the minimum of the bimodal metallicity distribution of GCs suggesting that it might be an excellent transitional system between metal-intermediate and metal-rich GCs. In spite of its metal-intermediate status, it is characterized by an extended horizontal branch and by a large number of RR Lyrae stars. We investigated abundances of a variety of species including light, $\alpha$-, Fe-peak, and neutron capture elements. We found a mean metallicity $[Fe/H]=-0.93 \pm 0.05$ dex, and a typical $\alpha$ -enrichment ($[\alpha/Fe] \approx 0.39$) that follows the trend of metal-poor and metal-intermediate GCs. The same outcome applies for light metals (Na, Al), Fe-peak (V, Cr, Mn, Fe, Co, Ni, Cu), $s$/$r$-process elements (Ba, Eu) and for the classical anti-correlation: Na-O and Mg-Al. The current findings further support the evidence that the chemical enrichment of NGC~6723 is in more line with metal-intermediate GCs and their lower metallicity counterparts, and it does not bring forward the prodrome of the metal-rich regime.
\end{abstract}


%%%%%%%%%%%%%%%%%%%%%%%%%%%%%%%%%%%%%%%%%%%%%%%%%%

%%%%%%%%%%%%%%%%% BODY OF PAPER %%%%%%%%%%%%%%%%%%

\section{Introduction} \label{sect:introduction}
Globular clusters (GCs) are important sources of information due to their extremely long lifetime, which makes them witnesses to primordial evolutionary stages of the Galaxy. While they were long assumed to be single stellar populations (i.e. coeval and chemically homogenous), it has been known for decades that they display peculiarities such as light element abundance variations \citep[e.g.][]{1971Obs....91..223O}, and varied horizontal branch (HB) morphologies that are often surprising given their metallicity \citep{1997ApJ...484L..25R}. More recently, with the aid of higher precision spectroscopy and photometry, a scenario of multiple stellar populations as the norm for GCs has emerged, with split sequences in color-magnitude diagrams (CMDs) and unexpected correlations between light elements that are yet to be understood \citep{2004MSAIS...5..105B,2007ApJ...661L..53P,2012A&ARv..20...50G,2018ARA&A..56...83B}. As different chemical species trace different astrophysical mechanisms (i.e. Type Ia Supernovae, winds from evolved stars), high resolution abundance estimates for a variety of elements and for many GCs are necessary for our understanding of these complex objects. Many remain poorly studied, especially regarding heavier species such as Fe-peak and neutron(n)-capture elements. Such is the case of NGC 6723.

Located at $R_{\odot} = 8.7$ kpc, it is an old \citep[$\approx 12.50$ Gyr,][]{2010ApJ...708..698D} GC that lies at a region of low foreground interstellar reddening, with $E(B-V) = 0.063$ mag \citep[][hereafter L14]{2014ApJS..210....6L}, making it an excellent candidate for the study of GCs in the densest regions of the Milky Way. It has been frequently assumed to belong to the Bulge due to its proximity to it at about $R_{GC} \approx 2.6$ kpc, however its dynamics point out to a inner Halo membership since the work of \citet{2003AJ....125.1373D}, a conclusion now confirmed by brand new proper motion estimates from {\it Gaia}\footnote{{\it Gaia} is a mission from the European Space Agency (ESA) mission (\url{https://www.cosmos.esa.int/gaia}), with data processed by the {\it Gaia} Data Processing and Analysis Consortium (DPAC, \url{https://www.cosmos.esa.int/web/gaia/dpac/consortium}). Funding for the DPAC has been provided by national institutions, in particular the institutions participating in the {\it Gaia} Multilateral Agreement.} \citep{2018arXiv180709775V,2018AstBu..73..162C}. 

Metallicity estimates for this object have covered a range from $[\ion{Fe}/\ion{H}] = -0.96 \pm 0.12$ (L14) to $-1.26 \pm 0.09$ \citep{1996ASPC...92..265F}, creating some uncertainty regarding its status as a metal-intermediate stellar system. The question of metallicity is particularly important because NGC 6723 presents an extended HB, where both the blue and red side of the RR Lyrae instability strip are populated \citep{1999A&AS..136..461A}. While the group of GCs that shares in this characteristic is very heterogeneous due to the so-called second parameter problem \citep{1967AJ.....72...70V,1994ApJ...423..248L,2009Ap&SS.320..261C,2014ApJ...785...21M}, extended HBs and the presence of many RR Lyrae stars are expected in metal-poor GCs. \citet{2015A&A...573A..92G}, with a sample of $47$ HB stars on both sides of the instability strip, determined that there is a wide range in color even among $43$ chemically homogeneous stars, which requires a considerable and not yet understood spread in mass loss along the red giant branch (RGB) phase and possibly $He$ abundance variations. 

While NGC 6723 has been studied at length photometrically, spectroscopic information about it remains sparse. The intermediate metallicity region represents a minimum in the bimodal metallicity distribution of Galactic GCs \citep[e.g.][]{1979ApJ...231L..19H,1981ARA&A..19..319F,1985ApJ...293..424Z,2006A&A...450..105B} and is poorly sampled, making the study of globulars in this regime especially important. In the present work, we have obtained abundance estimates for variety of chemical species with a sample of eleven RGB stars. Thus, we covered the main element families, namely the $\alpha$-elements (O, Mg, Si, Ca, Ti), Fe-peak elements (V, Cr, Mn, Fe, Co, Ni, Cu), products of rapid and slow n-capture (Ba and Eu) and the light odd-Z elements (Na, Al) that exhibit spreads in GCs. With this, we add high resolution spectroscopic estimates previously scarce or missing for this extended HB GC with conflicting metallicity estimates. In Sect.~\ref{sect:data}, we describe the observations and reduction of the data set, in Sect.~\ref{sect:analysis} we describe the derivation of atmospheric parameters and chemical abundances, which are discussed in Sect.~\ref{sect:results}. An overview regarding its inner Halo membership, HB morphology, multiple stellar populations, and general abundance pattern is given in Sect.~\ref{sect:overview}. Conclusions are presented in Sect.~\ref{sect:conclusions}.

\section{Observations and Data Reduction} \label{sect:data}

We analysed a total of eleven bright giant stars in NGC 6723. Four of which ($a1$ through $a4$) plus the reference star Arcturus had their spectra acquired in 2010 with the Magellan Inamori Kyocera Echelle \citep[MIKE,][]{2003SPIE.4841.1694B}, installed on the Magellan II "Clay", one of the twin 6.5 Magellan telescopes on Cerro Manqui at the Las Campanas Observatory, at a resolution of $R \approx 22000$. These data were reduced using the MIKE pipeline, with continuum normalization and Doppler shift correction being performed manually using the IRAF \footnote{\textsc{IRAF} is distributed by the National Optical Astronomy Observatory, which is operated by the Association of Universities for Research in Astronomy (AURA) under cooperative agreement with the National Science Foundation.} routines \emph{rvidlines}, \emph{dopcor}, and \emph{continuum}. The reduced and Doppler shift corrected spectra for the remaining seven ($b1$ through $b7$) were kindly provided by Dr. Rojas-Arriagada \citep[][hereafter RA16]{2016A&A...587A..95R}. They were obtained in 2010 with the Fibre-fed, Extended Range, \'{E}chelle Spectrograph (FEROS) at the MPG/ESO 2.2 m telescope on the La Silla Observatory, with $R \approx 48000$. All stars in the sample are cluster member candidates due to their radial velocities and proper motions (Table~\ref{tab:observations}).

\begin{figure}
	\includegraphics[width=\columnwidth]{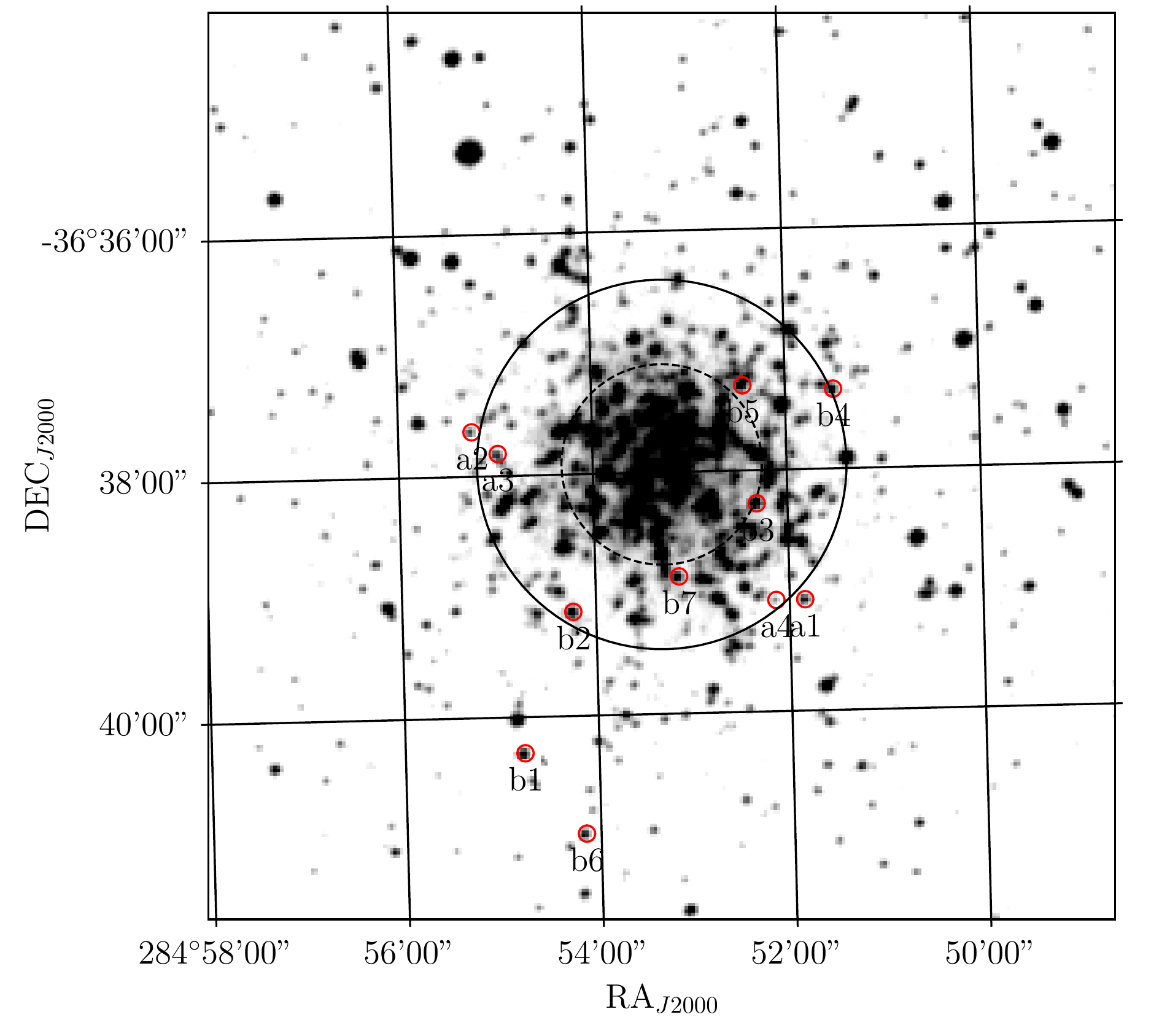}
    \caption{Finding chart for the analyzed stars, which are indicated by red circles. The dashed and full line black cirles represent the core and half-light radius, respectively \citep[][2010 edition]{1996AJ....112.1487H}. Image from the ESO Digitized Sky Survey.}
    \label{fig:findingchart}
\end{figure}

\begin{figure}
	\includegraphics[width=\columnwidth]{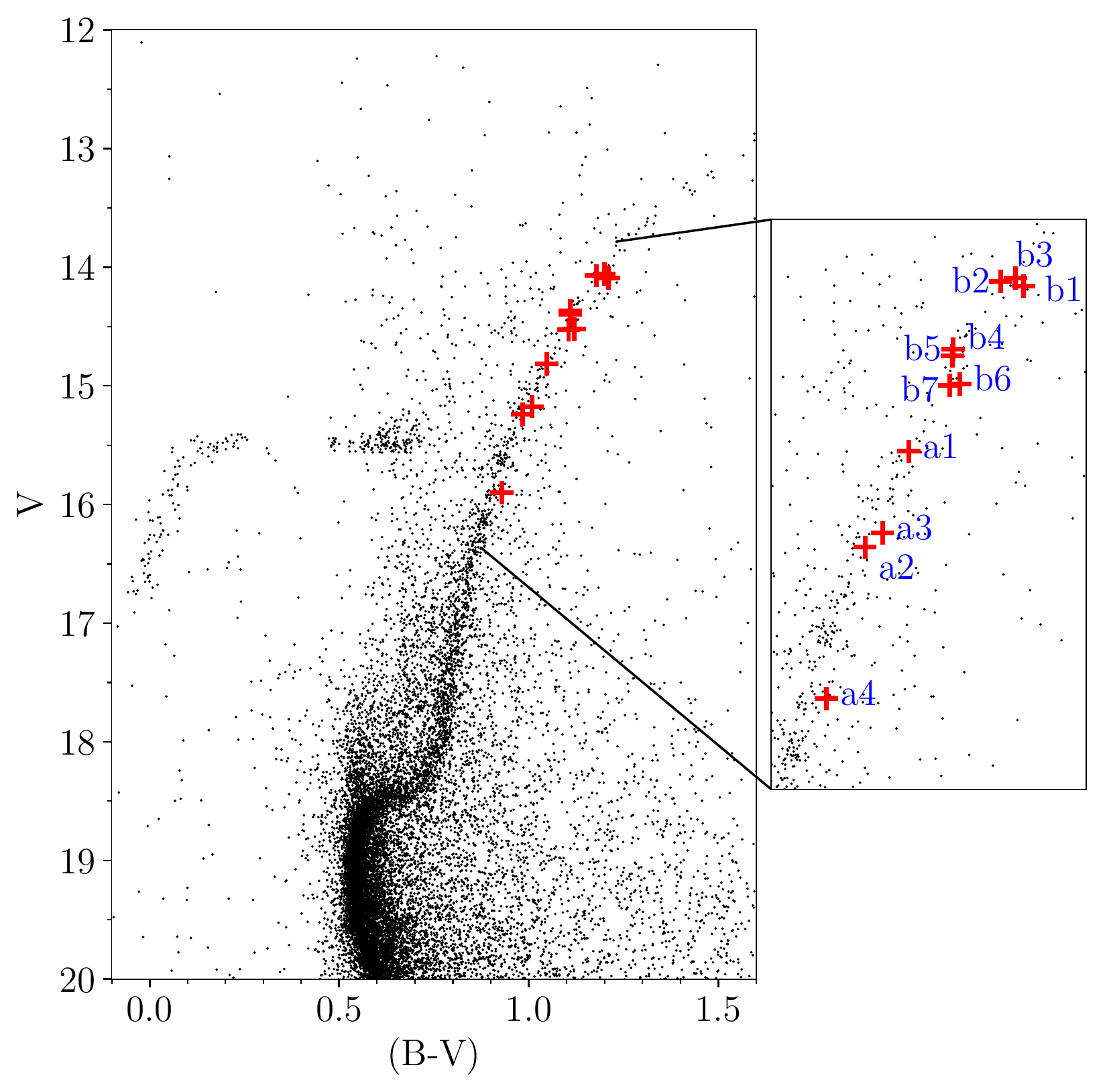}
    \caption{Color-magnitude diagram for NGC 6723 (L14). The analyzed stars are indicated by red crosses.}
    \label{fig:cmd}
\end{figure}

Signal-to-noise ratio (SNR) for the whole sample was computed with the IRAF task \emph{splot} using the mean value of three $2$ \r{A} intervals at the beginning, middle, and end of the wavelength range. The finding chart for the full sample is shown in Fig.~\ref{fig:findingchart} and Table~\ref{tab:observations} contains further details such as exposure times, SNR, and radial velocities. Photometric data in the $BV$ bands were kindly made available by Dr. Lee (L14), having been obtained from 2002 to 2012 using the $0.9$ m and the $1.0$ m telescopes from Cerro Tololo Inter-American Observatory (CTIO). See Fig.~\ref{fig:cmd} for the color-magnitude diagram (CMD). Additional photometric data were taken from the Two Micron All-Sky Survey \citep[2MASS,][]{2006AJ....131.1163S} All-Sky Point Source Catalog\footnote{The Two Micron All Sky Survey is a joint project of the University of Massachusetts and the Infrared Processing and Analysis Center/California Institute of Technology, funded by NASA and the National Science Foundation.}.

\begin{table*}
	\centering
	\caption{Identification and coordinates of the analyzed stars with the corresponding exposure time, airmass, seeing, SNR, and derived heliocentric radial velocities of their spectra. Proper motions were taken from the {\it Gaia} DR2 \citep{2016A&A...595A...1G,2018arXiv180409365G}.}
	\label{tab:observations}
	\begin{tabular}{cccccccrcc} 
		\hline
		Star ID & $\alpha_{J2000}$ & $\delta_{J2000}$ & Exp & Airmass & Seeing & SNR & $v_{hel}$ & $\mu_{\alpha}^*$ & $\mu_{\delta}$ \\
             & (deg)          & (deg)          &(s)&         & (arcsec)       & (\r{A}$^{-1}$)    & (km s$^{-1}$)&(mas yr$^{-1}$)&(mas yr$^{-1}$) \\
		\hline
		$a1$   & $284.865$ &	$-36.651$ & -- 		& $1.201$ & --     & $79$ & $-107.21$ & $1.11 \pm 0.05$ & $-2.54 \pm 0.05$ \\
		$a2$   & $284.922$ &	$-36.627$ & --		 & $1.010$ & --     & $70$ & $-100.64$ & $1.27 \pm 0.07$ & $-2.40 \pm 0.07$ \\
		$a3$   & $284.917$ &	$-36.630$ & --		 & $1.035$ & --     & $59$ & $-99.26 $ & $0.95 \pm 0.07$ & $-2.44 \pm 0.07$ \\
		$a4$   & $284.870$ &	$-36.651$ & --		& $1.064$ & --     & $48$ & $-101.12$ & $0.86 \pm 0.08$ & $-2.44 \pm 0.07$ \\
		$b1$ & $284.914$ &	$-36.672$ & $2700$    & $1.057$ & $0.93$ & $54$ & $-96.65$  & $1.18 \pm 0.06$ & $-2.42 \pm 0.06$ \\
		$b2$ & $284.905$ &	$-36.652$ & $5400$    & $1.160$ & $1.22$ & $56$ & $-90.01$  & $0.94 \pm 0.05$ & $-2.41 \pm 0.05$ \\
		$b3$ & $284.873$ &	$-36.638$ & $5400$    & $1.021$ & $1.11$ & $51$ & $-100.38$ & $1.01 \pm 0.06$ & $-2.45 \pm 0.04$ \\
		$b4$ & $284.859$ &	$-36.622$ & $4600$    & $1.450$ & $1.47$ & $38$ & $-99.78$  & $0.92 \pm 0.05$ & $-2.54 \pm 0.04$ \\
		$b5$ & $284.875$ &	$-36.622$ & $4500$    & $1.018$ & $1.56$ & $40$ & $-96.16$  & $1.08 \pm 0.08$ & $-2.20 \pm 0.07$ \\
		$b6$ & $284.904$ &	$-36.683$ & $4500$    & $1.211$ & $1.47$ & $21$ & $-94.08$  & $0.86 \pm 0.08$ & $-2.50 \pm 0.08$ \\
		$b7$ & $284.887$ &	$-36.648$ & $4500$    & $1.024$ & $0.84$ & $43$ & $-98.84$  & $0.96 \pm 0.06$ & $-2.54 \pm 0.05$ \\
		\hline
	\end{tabular}
\end{table*}

\section{Analysis} \label{sect:analysis}
\subsection{Photometric Parameters}

Adopting the typical value $R_V = 3.1$, $E(B-V) = 0.063$ mag (L14), and $A_{K_s} = 0.117$ \citep[][]{2017AstL...43..472G,1985ApJ...288..618R}, we un-reddened the $BV$ (L14) and $K_s$ (2MASS) magnitudes. With the intrinsic colors, we used two empirical color-temperature relations. The first employs the 2MASS bands \citep{2009A&A...497..497G} while the second uses the Johnson photometric system for $BV$ \citep{1999A&AS..140..261A,2001A&A...376.1039A}. Thus, we obtained the first photometric estimate for $T_{eff}$ for each star by taking a simple mean of the values derived from each color. The bolometric correction $BC(V)$ could then be obtained in order to compute the bolometric magnitude with the relation $M_{bol} = M_V + BC(V)$.

For mass estimates, we compared the observed CMD to a Dartmouth Stellar Evolution Database isochrone \citep{2008ApJS..178...89D}, considering the parameters derived by \citet{2010ApJ...708..698D} for NGC 6723. The differences in mass were negligible, with all stars presenting values close to $0.83 M_{\odot}$. With the intrinsic colors, bolometric correction, and mass estimates, it is possible to compute the photometric surface gravity $log(g)$ employing the classical relation:

\begin{equation}
log \left ( \frac{g_*}{g_{\odot}} \right ) = 4 log \left ( \frac{T_*}{T_{\odot}} \right ) + 0.4 \left (M^*_{bol} - M^{\odot}_{bol}\right ) + log \left ( \frac{M_*}{M_{\odot}} \right )
\label{eq.logg}
\end{equation}
and the Solar values $log(g_{\odot}) = 4.44$ dex, $T_{\odot} = 5780$ K, and $M_{bol}^{\odot} = 4.75$ dex. The mean of these estimates were used as first guesses for the spectroscopic parameters (Sect.~\ref{sect:spectroscopy}). Once spectroscopic metallicities were found, we used the color-temperature relations once again to derive final photometric values (Table~\ref{tab:atmosphere}). All chemical abundances were computed using exclusively the spectroscopic atmospheric parameters.

\begin{table*}
	\centering
	\caption{The infra-red (2MASS) and optical (L14) magnitudes, with their corresponding effective temperatures and surface gravities (see text for details), which were used as a first guess for the computation of the atmospheric parameters. The last four columns contain the final spectroscopic atmospheric parameters, which were the ones adopted in this work. Due to its lower SNR and likely spurious metallicity value, star $b6$ has not been considered in any computation of mean values and standard deviations.}
	\label{tab:atmosphere}
	\begin{tabular}{ccccccccccccc} 
		\hline
        \multicolumn{5}{c}{ } &  \multicolumn{2}{c}{Photometric}  &  \multicolumn{4}{c}{Spectroscopic} \\
		Star & $B$ & $V$ & $J$ & $K_s$ & $T_{eff(B-V)}$ & $log(g)_{(B-V)}$ & $T_{eff(V-K_s)}$ & $log(g)_{(V-K_s)}$ & $T_{eff}$ & $log(g)$ & $\xi_t$ & $[\ion{Fe}/\ion{H}]$\\
             & (mag) & (mag) & (mag) & (mag) & (K)		& (dex)              & (K)              & (dex)              & (K)       & (dex)    & (km s$^{-1}$) & (dex) \\
		\hline
		$a1$  & $15.864$ & $14.817$ & $12.728$ & $12.021$ & $4638$ & $1.90$ & $4586$ & $1.87$ & $4610$ & $2.01$ & $1.37$ & $-0.91$ \\
		$a2$  & $16.222$ & $15.238$ & $13.211$ & $12.564$ & $4730$ & $2.12$ & $4686$ & $2.09$ & $4750$ & $2.28$ & $1.54$ & $-0.98$ \\
		$a3$  & $16.184$ & $15.175$ & $13.159$ & $12.488$ & $4705$ & $2.08$ & $4675$ & $2.06$ & $4790$ & $2.57$ & $1.55$ & $-0.88$ \\
		$a4$  & $16.830$ & $15.901$ & $14.028$ & $13.379$ & $4826$ & $2.44$ & $4819$ & $2.44$ & $4900$ & $2.96$ & $1.67$ & $-0.97$ \\
		$b1$  & $15.302$ & $14.091$ & $11.788$ & $10.987$ & $4395$ & $1.45$ & $4358$ & $1.43$ & $4470$ & $1.70$ & $1.53$ & $-0.83$ \\
		$b2$  & $15.250$ & $14.072$ & $11.806$ & $11.020$ & $4428$ & $1.47$ & $4394$ & $1.45$ & $4490$ & $1.52$ & $1.71$ & $-0.97$ \\
		$b3$  & $15.256$ & $14.057$ & $11.820$ & $11.041$ & $4396$ & $1.44$ & $4420$ & $1.46$ & $4330$ & $1.10$ & $1.42$ & $-0.99$ \\
		$b4$  & $15.479$ & $14.369$ & $12.263$ & $11.522$ & $4530$ & $1.65$ & $4546$ & $1.66$ & $4510$ & $1.74$ & $1.55$ & $-0.97$ \\
		$b5$  & $15.507$ & $14.398$ & $12.259$ & $11.537$ & $4535$ & $1.67$ & $4535$ & $1.67$ & $4570$ & $1.86$ & $1.68$ & $-0.94$ \\
		$b6$  & $15.641$ & $14.521$ & $12.329$ & $11.594$ & $4553$ & $1.73$ & $4485$ & $1.69$ & $4660$ & $1.69$ & $1.85$ & $-0.70$ \\
		$b7$  & $15.632$ & $14.527$ & $12.368$ & $11.659$ & $4546$ & $1.73$ & $4530$ & $1.72$ & $4445$ & $1.39$ & $1.32$ & $-0.91$ \\
		\hline
	\end{tabular}
\end{table*}

\subsection{Line List and Spectroscopic Parameters} \label{sect:spectroscopy}
The list of atomic lines for all species without hyperfine structure (HFS) was compiled from \citet{2010A&A...513A..35A} and \citet{2014MNRAS.439.2638Y}. We have measured equivalent widths (EW) manually using the \textsc{IRAF} task \emph{splot} for all species from those lists except for O, which was treated the same way as the species with HFS described ahead. Only lines with EW between $15$ and $150$ m\r{A} were used to avoid both continuum noise and lines that deviated from the Gaussian profile required for our analysis. For Sc, V, Co, Cu, and Ba, all elements with HFS, we have adopted the line lists from \citet{2000AJ....120.2513P}. For Mn, we have employed the line list from \citet{2013A&A...559A...5B}. For the Eu line, we employed the method described in the Appendix of \citet{2000AJ....120.2513P} and the transition parameters from \citet{2001ApJ...563.1075L} to compute the HFS values. Solar abundance values from \citet{2009ARA&A..47..481A} were adopted for all species.

In order to compute the spectroscopic atmospheric parameters (effective temperature $T_{eff}$, surface gravity $log(g)$, microturbulent velocity $\xi_t$, and metallicity $[\ion{Fe}/\ion{H}]$), we performed a LTE 1D plane-paralel analysis using the \emph{abfind} task from the \textsc{MOOG} code \citep[][2014 version]{1973ApJ...184..839S} and an interpolated grid of Kurucz atmospheric models built with the \textsc{ATLAS9} code \citep{1997A&A...318..841C} without convective overshooting and a Solar chemical mixture. The procedure followed the traditional methods, where the $Fe$ line-by-line abundances are plotted against their respective excitation potentials ($\chi$) and reduced equivalent widths ($log(EW/\lambda)$), changing the amospheric models iteratively to eliminate trends on both accounts. The spectroscopic $T_{eff}$ is found when excitation equilibrium is achieved, which happens when the linear fit of \ion{Fe}{I} abundances versus $\chi$ has a null slope. The microturbulence velocity $\xi_t$ is found when vanishing residuals are obtained for \ion{Fe}{I} abundances versus $log(EW/\lambda)$. When the \ion{Fe}{I} and \ion{Fe}{II} abundances are equal, ionization equilibrium is reached, and the spectroscopic surface gravity $log(g)$ is found. Finally, the adopted metallicity $[\ion{Fe}/\ion{H}]$ is the one found when all these criteria are satisfied, with the metallicity found by MOOG and the input metallicity being equal. 

\begin{figure}
	\includegraphics[width=\columnwidth]{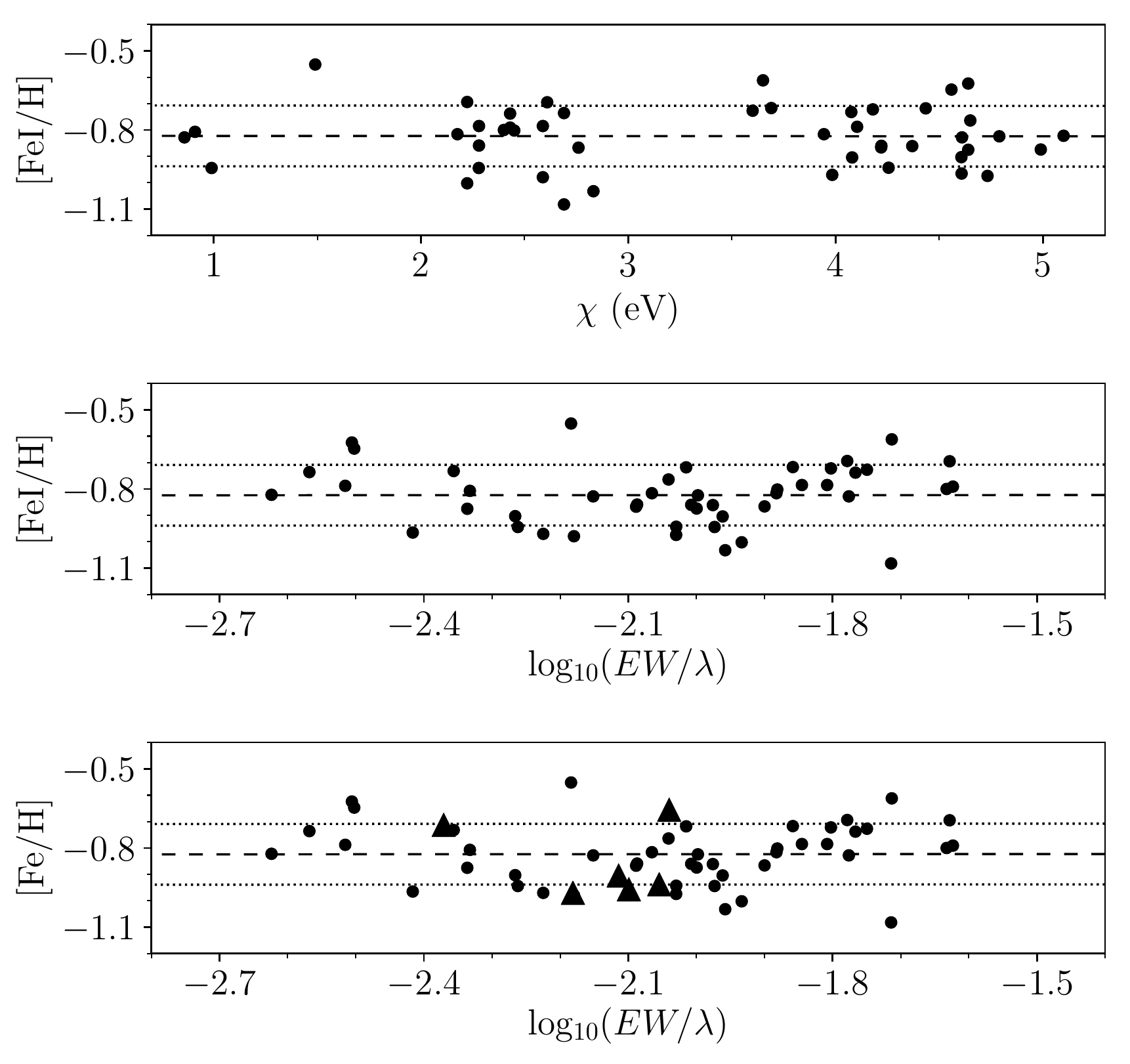}
    \caption{Final result of iterative process to compute the spectroscopic atmospheric parameters for star $a1$. Circles and triangles represent, respectively, FeI and FeII lines. Any trends have been eliminated to slopes equal to or smaller than $\pm 0.001$. See text for details.}
    \label{fig:correlations}
\end{figure}

Since changes in different atmospheric parameters have varying levels of influence on all the aforementioned trends, the process must be repeated with increasingly finer adjustements until the three trends are eliminated and the final atmospheric parameters are found. We have considered the final parameters to be converged only when trends were equal to or smaller than $\pm 0.001$ (Fig. \ref{fig:correlations}). Therefore, our $[\ion{Fe}{I}/H]$ and $[\ion{Fe}{II}/\ion{H}]$ values are the same up to the third decimal position. The final parameters are presented in Table~\ref{tab:atmosphere}. For the reference star Arcturus, the derived values are $T_{eff} = 4365$ K, $log(g) = 1.4$ dex, $[\ion{Fe}/\ion{H}] =  -0.5$ dex, and $\xi_t = 1.55$ km s$^{-1}$, in good agreement with \citet{2011ApJ...743..135R}.

\subsection{Abundances}
\begin{figure}
	\includegraphics[width=\columnwidth]{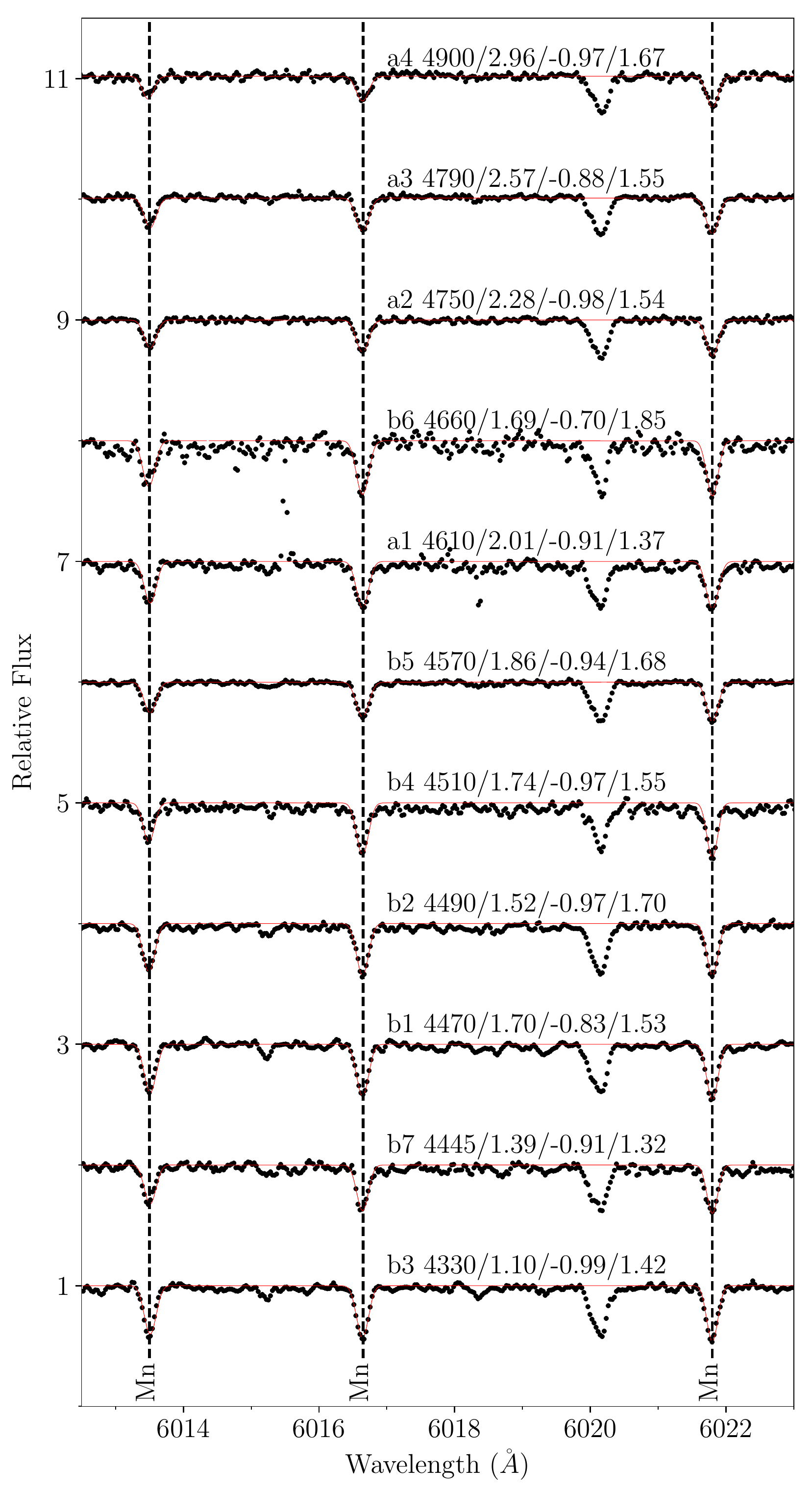}
    \caption{\emph{Black dots:} portion of the observed spectra for the whole sample, ordered by temperature. Each star is indicated, alongside its atmospheric parameters $T_{eff}$/$log(g)$/$[\ion{Fe}/\ion{H}]$/$\xi_t$ in $K$, dex, dex, and kms$^{-1}$ units, respectively. \emph{Red line:} the corresponding modelled manganese lines.}
    \label{fig:spectra_example}
\end{figure}

Once the final atmospheres were computed, abundances for all other elements without HFS can be derived line-by-line using the same \textsc{MOOG} task \emph{abfind}. As a first step, we computed Tukey's Biweight, a robust estimation of central tendency, for each species in each star. Then we eliminated lines that deviated more than $2.5 \sigma_{lines}$ from the first computation for the species in question and used the remaining lines to arrive at the final values. For O, Sc, V, Mn, Co, Cu, Ba, and Eu, we have employed the \textsc{MOOG} task \emph{synth} instead, and manually tweaked the $[X/Fe]$ abundance of each line until the synthetic profile matched the observed profile (Fig.~\ref{fig:spectra_example}). 

\subsection{Uncertainties}

Abundance values are sensitive to all atmospheric parameters depending on the species and line in question. To estimate their uncertainties, we have varied the atmospheric parameters one at a time by the typical, albeit conservative, values of $\Delta T_{eff} = \pm 150$ K, $\Delta log(g) = \pm 0.3$ dex, $\Delta [\ion{Fe}/\ion{H}] =  \pm 0.1$ dex, and $\Delta \xi_t = \pm 0.2$ km s$^{-1}$, and then computed the abundances for all species for each of these possibilities for two stars, one for each instrument employed in the acquisition of our sample. Thus, each line for each star has a corresponding $\sigma_{T_{eff}}$ which refers to its response to $\Delta T_{eff}$, and similarly for the other parameters. We propagate these uncertanties in quadrature to compute the uncertainty 
\begin{equation}
\sigma = \sqrt{\sigma_{T_{eff}}^2 + \sigma_{log(g)}^2 + \sigma_{[\ion{Fe}/\ion{H}]}^2 + \sigma_{\xi_t}^2}
\end{equation}
of each species per star (Table~\ref{tab:uncertainties}). Finally, we adopted as the mean $\sigma$ per species considering these two stars. 

Another important uncertainty estimate is the line-by-line variation for the same species within the same star, which is due to several sources such as continuum placement, instrumental noise, and uncertainties in atomic parameters. We have opted not to include $\sigma_{lines}$ in the computation of $\sigma$ as some elements have only one line, and that would result in an artificially lower $\sigma$. However, due to the importance of this source of uncertainty, $\sigma_{lines}$ is presented as well in Table~\ref{tab:uncertainties}. We have employed the additional measurement $\sigma_*$, which is the standard deviation that arises from the star-by-star scatter in the mean of each species and can serve as a diagnostic of whether $\sigma$ is under- or overestimated for species for which no intrinsic spread is expected. Note that due to the lower SNR of the spectrum for star $b6$, it has been removed from all computations of mean values and stardard deviations.

The forbidden O line at $6300$ \r{A} is robust against non-LTE effects and the blended Ni is weak at sub-solar metallicities, becoming negligible under $[\ion{Fe}/\ion{H}] = -1$ \citep{2001ApJ...556L..63A}, a fact which we verified in our spectra. Employing atmospheric models without $[\alpha/Fe]$ enhancement can result in artificially lower abundances of $[O/Fe]$ as it is a dominant species \citep[e.g.][]{1966IAUS...26..112D,1993ApJ...414..580S}. \citet{2008A&A...484L..21M} report that MARCS models with $[\alpha/Fe] = 0.2$ result in $[O/Fe]$ about $0.1$ dex higher than models without $\alpha$-enhancement. The Ni line is significant in the Sun, and our adopted Solar values already take this into account. Non-LTE effects have been studied for Fe-peak elements and will be discussed further in Sect.~\ref{sect:fe_peak_elements}.

\begin{table}
	\centering
	\caption{Sensitivity to typical uncertainties in atmospheric parameters and standard deviation between lines of the same species for stars $b4$ and $a1$. See text for details. Iron lines represent the $[\ion{Fe}{I}/H]$ and $[\ion{Fe}{II}/\ion{H}]$ values, while all other species are the $[X/Fe]$ ratios. As the spectrum for star $b4$ did not have serviceable Co lines, values for star $b1$ were used instead.}
	\label{tab:uncertainties}
	\begin{tabular}{ccccccr} 
		\hline
		Species & $\sigma_{Teff}$ & $\sigma_{[\ion{Fe}/\ion{H}]}$ & $\sigma_{log(g)}$ & $\sigma_{\xi_t}$ & $\sigma$ & $\sigma_{lines}$ \\
		\hline
        \multicolumn{7}{c}{Star $a1$} \\
        \hline
		\ion{Fe}{I}  & $0.14$ &	$0.00$ & $0.01$ & $0.08$ & $0.16$ & $0.12$  \\
		\ion{Fe}{II} & $0.14$ &	$0.04$ & $0.15$ & $0.06$ & $0.21$ & $0.15$  \\
 		\ion{O}{I}   & --	   & 	--	   & --	    & --	 & --	  & --      \\
 		\ion{Na}{I}  & $0.13$ &	$0.01$ & $0.01$ & $0.04$ & $0.14$ & $0.15$  \\
 		\ion{Mg}{I}  & $0.09$ &	$0.00$ & $0.02$ & $0.03$ & $0.10$ & $0.15$  \\
 		\ion{Al}{I}  & $0.11$ &	$0.01$ & $0.01$ & $0.01$ & $0.11$ & $0.03$  \\
 		\ion{Si}{I}  & $0.03$ &	$0.01$ & $0.05$ & $0.03$ & $0.07$ & $0.09$  \\
 		\ion{Ca}{I}  & $0.16$ &	$0.01$ & $0.03$ & $0.10$ & $0.19$ & $0.09$  \\
 		\ion{Sc}{II} & $0.01$ &	$0.07$ & $0.12$ & $0.05$ & $0.14$ & $0.09$  \\
 		\ion{Ti}{I}  & $0.22$ &	$0.01$ & $0.01$ & $0.04$ & $0.23$ & $0.07$  \\
 		\ion{V}{I}   & $0.28$ &	$0.10$ & $0.03$ & $0.01$ & $0.29$ & $0.11$  \\
 		\ion{Cr}{I}  & $0.14$ &	$0.01$ & $0.01$ & $0.02$ & $0.14$ & --      \\
 		\ion{Mn}{I}  & $0.18$ &	$0.10$ & $0.05$ & $0.04$ & $0.22$ & $0.01$  \\
 		\ion{Co}{I}  & $0.16$ &	$0.09$ & $0.03$ & $0.01$ & $0.19$ & $0.07$  \\
 		\ion{Ni}{I}  & $0.07$ &	$0.02$ & $0.06$ & $0.08$ & $0.13$ & $0.12$  \\
 		\ion{Cu}{I}  & $0.16$ &	$0.09$ & $0.03$ & $0.01$ & $0.19$ & --      \\
 		\ion{Ba}{II} & $0.05$ &	$0.05$ & $0.08$ & $0.23$ & $0.25$ & $0.02$  \\ 
 		\ion{Eu}{II} & $0.00$ &	$0.05$ & $0.13$ & $0.03$ & $0.14$ & --      \\
		\hline
        \multicolumn{7}{c}{Star $b4$} \\
        \hline
		\ion{Fe}{I}  & $0.13$ &	$0.00$ & $0.02$ & $0.08$ & $0.15$ & $0.18$ \\
		\ion{Fe}{II} & $0.16$ &	$0.05$ & $0.16$ & $0.06$ & $0.24$ & $0.13$ \\
 		\ion{O}{I}   & $0.05$ &	$0.05$ & $0.13$ & $0.00$ & $0.14$ & --     \\
 		\ion{Na}{I}  & $0.14$ &	$0.01$ & $0.01$ & $0.03$ & $0.14$ & $0.12$ \\
 		\ion{Mg}{I}  & $0.09$ &	$0.00$ & $0.00$ & $0.04$ & $0.10$ & $0.03$ \\
 		\ion{Al}{I}  & $0.12$ &	$0.00$ & $0.01$ & $0.01$ & $0.12$ & $0.11$ \\
 		\ion{Si}{I}  & $0.04$ &	$0.02$ & $0.05$ & $0.03$ & $0.08$ & $0.13$ \\
 		\ion{Ca}{I}  & $0.17$ &	$0.00$ & $0.03$ & $0.10$ & $0.20$ & $0.11$ \\
 		\ion{Sc}{II} & $0.01$ &	$0.07$ & $0.12$ & $0.06$ & $0.15$ & $0.05$ \\
 		\ion{Ti}{I}  & $0.24$ &	$0.01$ & $0.01$ & $0.04$ & $0.24$ & $0.14$ \\
 		\ion{V}{I}   & $0.28$ &	$0.10$ & $0.02$ & $0.00$ & $0.30$ & $0.05$ \\
 		\ion{Cr}{I}  & $0.15$ &	$0.01$ & $0.01$ & $0.02$ & $0.15$ & --     \\
 		\ion{Mn}{I}  & $0.15$ &	$0.10$ & $0.04$ & $0.05$ & $0.19$ & $0.04$ \\
 		\ion{Co}{I}  & $0.13$ &	$0.09$ & $0.05$ & $0.00$ & $0.17$ & $0.00$ \\
 		\ion{Ni}{I}  & $0.09$ &	$0.01$ & $0.05$ & $0.06$ & $0.12$ & $0.13$ \\
 		\ion{Cu}{I}  & $0.15$ &	$0.08$ & $0.05$ & $0.00$ & $0.18$ & --     \\
 		\ion{Ba}{II} & $0.07$ &	$0.06$ & $0.09$ & $0.21$ & $0.24$ & $0.00$ \\ 
 		\ion{Eu}{II} & $0.03$ &	$0.08$ & $0.13$ & $0.01$ & $0.15$ & --     \\
		\hline
	\end{tabular}
\end{table}

\section{Results and discussion} \label{sect:results}

\subsection{Metallicity} \label{sect:metallicity}
We found a mean metallicity $[\ion{Fe}/\ion{H}] = -0.93$ with $\sigma_* = 0.05$ dex when excluding the lower SNR and outlier star $b6$ from the sample. This is in agreement with previous works that employ varied methods but higher than the traditional $[\ion{Fe}/\ion{H}] \approx -1.20$ adopted for this object (Table~\ref{tab:metallicity_compare}). A deviation greater than uncertainties is found in the comparison to the photometric values of L14 in the \citet[][hereafter ZW84]{1984ApJS...55...45Z} calibration ($[\ion{Fe}/\ion{H}]_{ZW} = -1.23 \pm 0.11$ dex). However, this value corresponds to $[\ion{Fe}/\ion{H}]_{CG} = -1.03$ in the metallicity scale of \citet[][hereafter CG97]{1997A&AS..121...95C}, which is based on high dispersion observations. As their study was focused on RR Lyrae variables, L14 also employed the empirical relation between the metallicity of RRab stars and the Fourier decomposition parameters of their light curves \citep{1996A&A...312..111J}, finding $[\ion{Fe}/\ion{H}]_{JK96} = -0.94 \pm 0.12$ dex for their sample of $10$ regular RRab variables. The same work placed NGC 6723 firmly among the most metal-rich Oosterhoff type I clusters, with a mean period of RRab stars $<P_{ab}> = 0.541 \pm 0.066$ day and a number ratio of RRc to total number of RRLs $n_c / n_{ab+c} = 0.167$.

\citet{1997PASP..109..883R} obtained low resolution spectra ($\approx 4$ \r{A} in the $7250 - 9000$ \r{A} region) with SNR ranging from $15$ to $120$ for $818$ stars in $52$ GCs. They used the equivalent widths of the $Ca II$ triplet lines, which are some of the strongest features in this region of the spectra of evolved stars, to determine the calcium index $W'$ as a mean among all analyzed stars in a given GC. Transforming their $W'$ measurements into an iron abundance by using the ZW84 and GC97 photometric calibrations, they found $[\ion{Fe}/\ion{H}]_{ZW} = -1.09 \pm 0.14$ dex and $[\ion{Fe}/\ion{H}]_{CG} = -0.96 \pm 0.04$ dex for NGC 6723, and observed that $W'$ scales linearly to $[\ion{Fe}/\ion{H}]_{CG}$, but nonlinearly with $[\ion{Fe}/\ion{H}]_{ZW}$.

\citet{2003PASP..115..143K} presented a GC metallicity scale using $\ion{Fe}{II}$ lines in the spectra of $149$ giant stars in $11$ GCs. They found that their $[\ion{Fe}{II}/\ion{H}]_{KI}$ correlates linearly with $W'$ as defined by \citet{1997PASP..109..883R}, with coefficients that depend on the adopted atmospheric model. From this relationship, they derived $[\ion{Fe}{II}/\ion{H}]_{KI} = -1.11$ and $-1.03$ for MARCS and Kurucz models (the latter with and without convective overshooting), respectively, for NGC 6723.

There is a small number of direct high dispersion measurements of iron lines in NGC 6723. \citet{1996ASPC...92..265F} observed three red giants with the CTIO $4$ m echelle spectrograph with $R \approx 33000$, finding the mean value $[\ion{Fe}/\ion{H}] = -1.26 \pm 0.09$ dex. They used ATLAS9 model atmospheres and $39$ \ion{Fe}{I} lines. The authors called this a preliminary result and, unfortunately, the work appears to have been discontinued. It is unclear if $\ion{Fe}{II}$ lines were employed, as the line list was not published. As the photometric temperatures are from $200$ to $320$ K higher than spectroscopic estimates, there may be a systematic error in the effective temperatures that would, in turn, result in artificially lower metallicities.

\citet{2015A&A...573A..92G} analyzed $51$ HB stars in NGC 6723, finding an average of $[\ion{Fe}/\ion{H}] = -1.22 \pm 0.08$ dex for $30$ stars in the red side of the instability strip. The authors used moderatedly high resolution spectra (R $\approx 18700$) to determine chemical abundances for multiple species, however iron lines were not used to determine the atmospheric parameters. Indeed, $T_{eff}$ and surface gravities were determined photometrically from $(B-V)$ and $(V-K)$ colors, while microturbulence velocity $\xi_{t}$ for red and blue HB stars and metal abundance $[A/H]$ were fixed at $1.4$ km s$^{-1}$, $3.0$ km s$^{-1}$ and $-1.25$ dex, respectively.

\begin{figure}
	\includegraphics[width=\columnwidth]{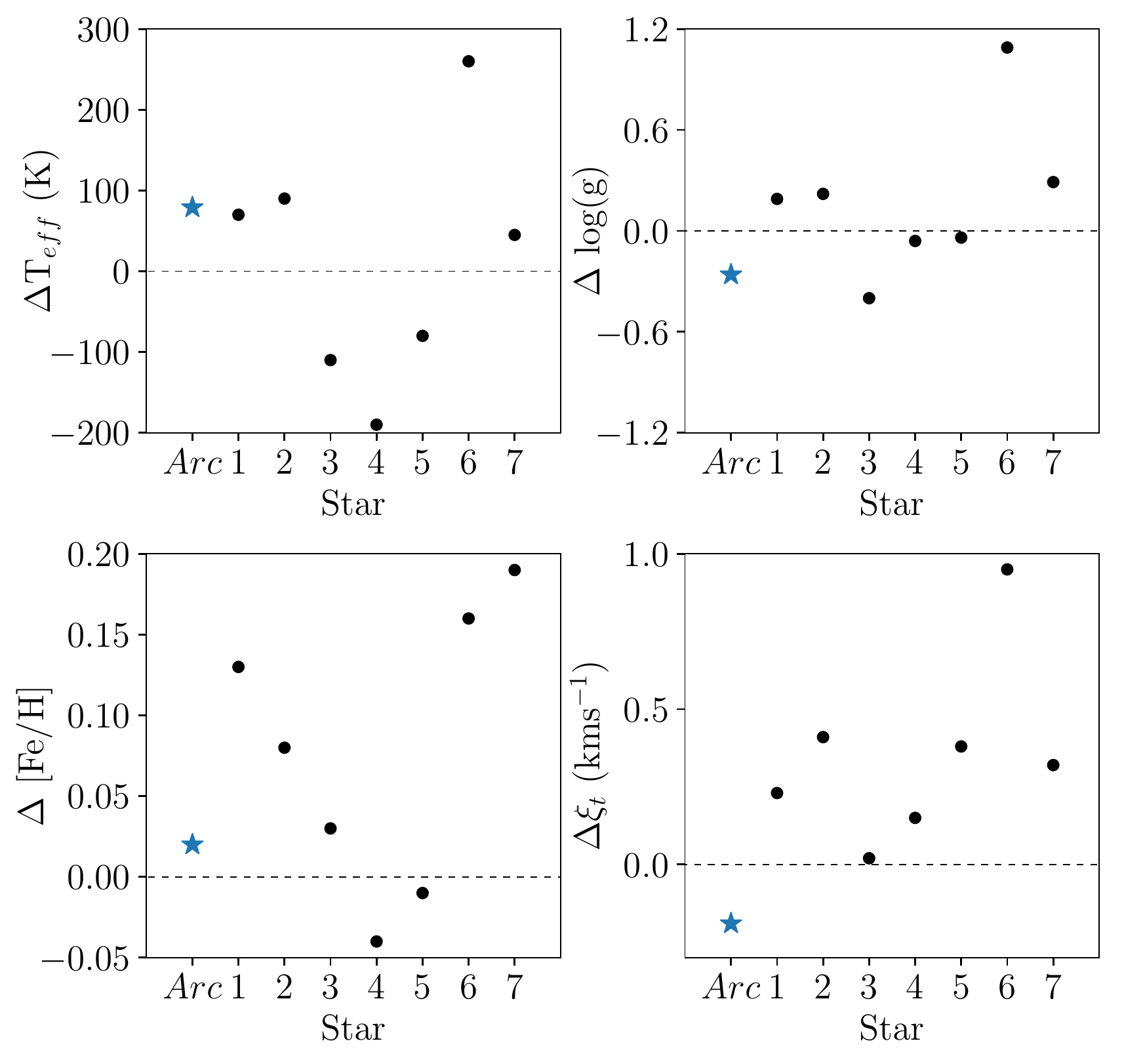}
    \caption{Difference between atmospheric parameters in this work and those of RA16. The difference between our values and those of \citet{2011ApJ...743..135R} for Arcturus is also shown as a quality test for our line list and methodology. In the abscissa, values for Arcturus are represented as $Arc$, and for stars $b1$ to $b7$ as $1$ through $7$. Our microturbulence velocities are consistently higher than those of RA16, while other parameters do not exhibit trends.}
    \label{fig:rojas_compare}
\end{figure}

RA16 performed an LTE analysis with a MARCS grid of spherical models \citep{2003ASPC..288..331G}, and found an average of $-0.98 \pm 0.08$ dex by using the weighted mean of iron lines according to their $EW$ error to compute $[\ion{Fe}/\ion{H}]$. They have adopted model-predicted $EW_p$ of \ion{Fe}{I} lines in the determination of microturbulent velocities. The GC mean values for metallicity and other elemental abundances are in good agreement despite sometimes significant disagreement for individual stars, without any significant correlation (Fig.~\ref{fig:rojas_compare}). Star $b6$ presents the largest deviation, which can be explained by its significantly lower SNR.

\begin{table}
	\centering
	\caption{Literature values for NGC 6723. Sources: [1] this work, [2] L14 , [3] \citet{1997PASP..109..883R}, [4] RA16, [5] \citet{1987ApJS...65...83M}, [6] \citet{2010ApJ...708..698D}, [7] \citet{2003PASP..115..143K}, [8] \citet{2015A&A...573A..92G}, [9] \citet{1996ASPC...92..265F}.}
	\label{tab:metallicity_compare}
	\begin{tabular}{lll} 
		\hline
		$[\ion{Fe}/\ion{H}]$ & Method & Source \\
		\hline
        $-0.93 \pm 0.05$ & Spectroscopy, RGB stars     & [1] \\
        $-0.94 \pm 0.12$ & Photometry, RRab variables  & [2] \\
        $-0.96 \pm 0.04$ & $W'$ index, CG97 calibration  & [3] \\
        $-0.98 \pm 0.08$ & Spectroscopy, RGB stars    & [4] \\
        $-1.00 \pm 0.21$           & Photometry, ZW84 calibration  & [5] \\ 
        $-1.00$          & Photometry, isochrone fitting & [6] \\
        $-1.03$          & \ion{Fe}{II} lines, Kurucz models & [7] \\		
        $-1.09 \pm 0.14$ & $W'$ index, ZW84 calibration  & [3] \\
        $-1.11$          & \ion{Fe}{II} lines, MARCS models  & [7] \\
        $-1.22 \pm 0.08$ & Spectroscopy, HB stars      & [8]\\
        $-1.23 \pm 0.11$ & Photometry, ZW84 calibration  & [2] \\
        $-1.26 \pm 0.09$ & Spectroscopy, RGB stars     & [9] \\
		\hline
	\end{tabular}
\end{table}

\subsection{$\alpha$-elements} \label{sect:alpha_elements}
Different $\alpha$-elements trace different enrichment mechanisms \citep[see, for example,][and references therein]{2016PASA...33...40M}. Observing them in a variety of sites is necessary to validate theoretical models and tweak their details, constraining important parameters such as the timescale of chemical enrichment in a stellar system. NGC 6723 displays considerable $\alpha$-element enhancement, a clear signature of the fast enrichment provided by supernovae (SNe) II events. This is expected in old stellar systems such as GCs, as they were formed before SNe Ia could contribute significantly with iron. 

\begin{figure}
	\includegraphics[width=\columnwidth]{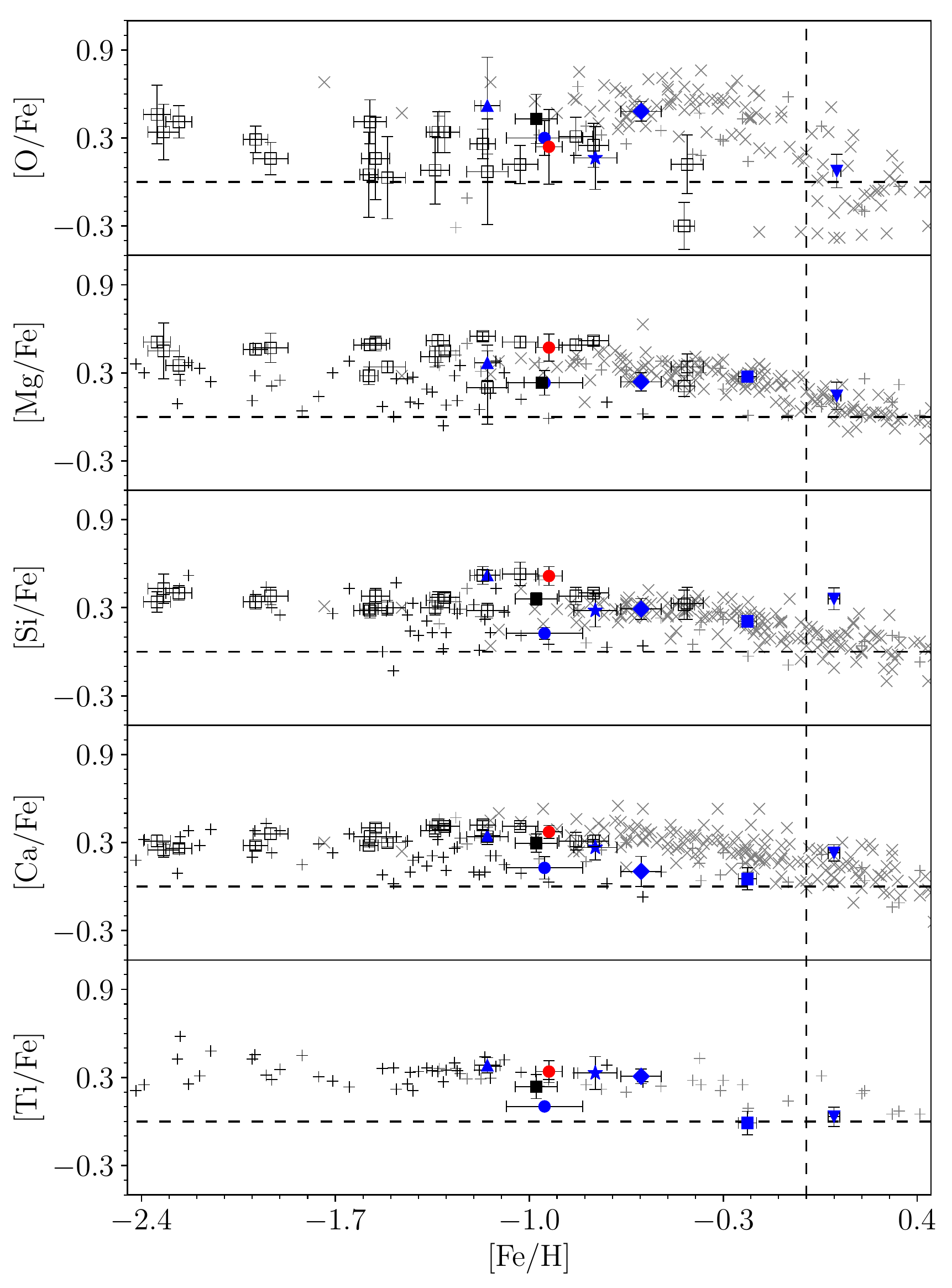}
    \caption{\emph{Red cirles:} this work. \emph{Black filled square:} stars 1 to 7 as computed by RA16. \emph{Blue star:} 47 Tucanae \citep{2014ApJ...780...94C}. \emph{Blue triangle:} NGC 6266 \citep[M62,][]{2014MNRAS.439.2638Y}. \emph{Blue upside down triangle:} NGC 6528 \citep{2001AJ....122.1469C}. \emph{Blue circle:} NGC 6522 \citep{2014A&A...570A..76B}. \emph{Blue square:} NGC 6553 \citep{2006A&A...460..269A}. \emph{Blue diamond:} NGC 6366 \citep{2018MNRAS.476..690P}. \emph{Black empty squares:} GCs from \citet[][Ca]{2010ApJ...712L..21C} and \citet[][remaining species]{2009A&A...505..139C}.\emph{Grey crosses and 'x':} Bulge field stars from \citet{2010A&A...513A..35A} and \citet{2014AJ....148...67J}, respectively. \emph{Black crosses:} inner Halo field stars \citep{2012ApJ...753...64I}. The error bars represent the standard deviation in the star-to-star scatter. No corrections for different adopted Solar abundance values were performed.}
    \label{fig:alpha_elements}
\end{figure}

The mean abundance lies at $[\alpha/Fe] = 0.39$, when considering the traditional group O, Mg, Si, Ca, and Ti. This value remains about the same when hydrostatic and explosive $\alpha$-elements are separated, with $[(O+Mg)/Fe] = 0.36 $ and $[(Si+Ca+Ti)/Fe] = 0.41 $. Mean values are in good agreement with GCs of similar metallicity (Fig.~\ref{fig:alpha_elements}), with small star-to-star spread with the exception of O in likely second generation star $b3$ (see Sect.~\ref{sect:oddz_elements} for details).

\subsection{Light odd-Z elements} \label{sect:oddz_elements}

Light elements have long been known to vary within GCs, even when an iron spread is absent. This appears to be evidence of two or more star formation episodes within the same GC, resulting in the presence of multiple stellar populations with varying abundances of certain species. Considering samples of random stars in a given GC, a Na-O anticorrelation appears to be ubiquitous, and perhaps a Mg-Al anticorrelation may be a feature of massive and metal-poor GCs \citep[e.g.][]{2004ARA&A..42..385G,2012A&ARv..20...50G}. The origin of these peculiar chemical patterns is still a matter of debate, with pollution from asymptotic giant branch (AGB) stars being a favored scenario. However, none of the current hypotheses on the nature of the polluters and the mechanism behind the detected number ratio of so-called "first" and "second generation" (FG and SG, respectively) stars can fully explain observational constraints \citep{2018ARA&A..56...83B,2018A&A...616A.168L}, therefore it is fundamental to explore the chemistry of as many GCs as possible in order to tweak and perhaps discard possibilities.

\begin{figure}
	\includegraphics[width=\columnwidth]{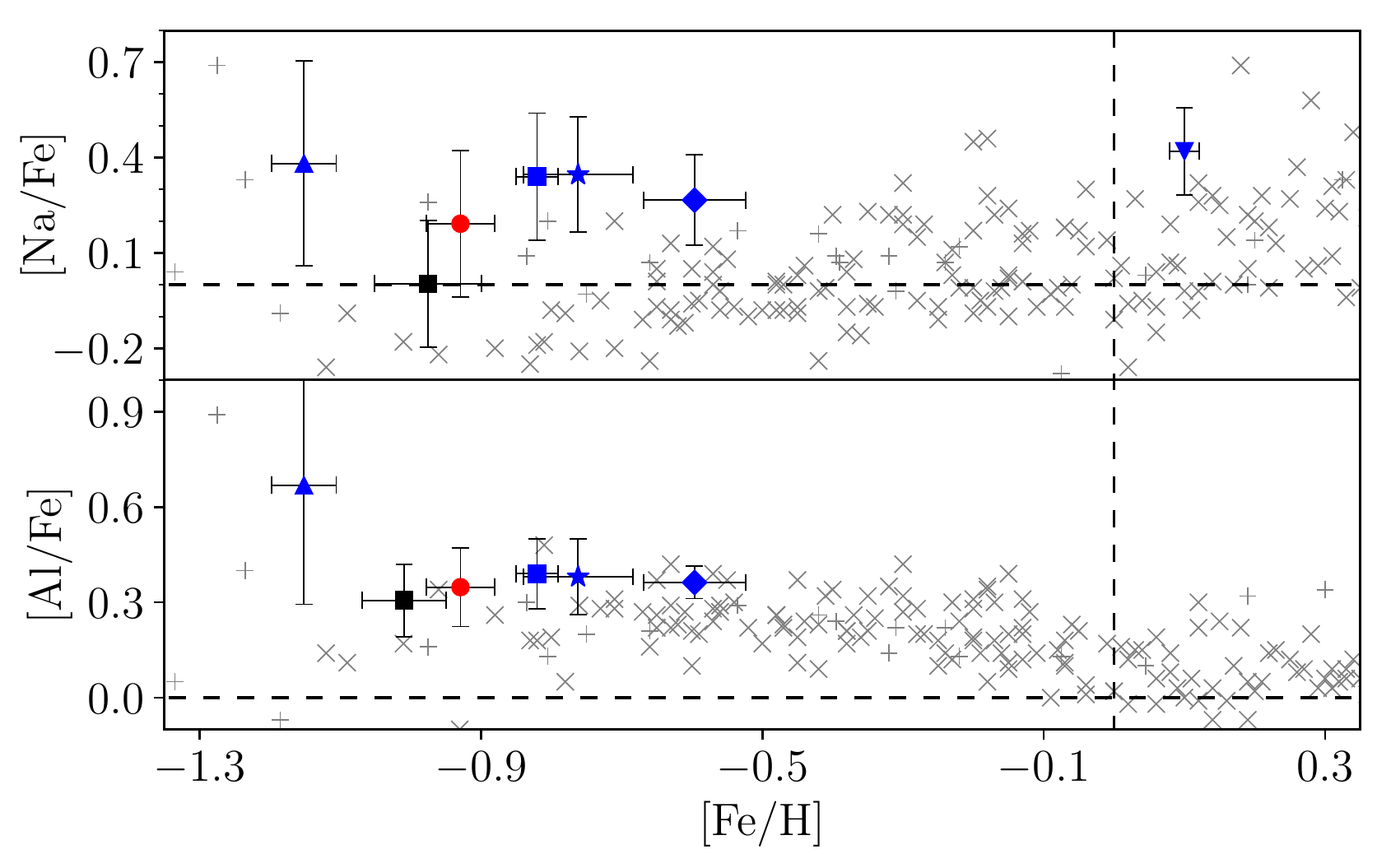}
    \caption{\emph{Red cirles:} this work. \emph{Black square:} stars 1 to 7 as computed by RA16. \emph{Blue star:} 47 Tucanae \citep{2014ApJ...780...94C}. \emph{Blue square:} NGC 6838 \citep{2015ApJ...800....3C}. \emph{Blue triangle:} NGC 6266 \citep[M62,][]{2014MNRAS.439.2638Y}. \emph{Blue upside down triangle:} NGC 6528 \citep{2001AJ....122.1469C}. \emph{Blue diamond:} NGC 6366 \citep{2018MNRAS.476..690P}. \emph{Grey crosses and 'x':} Bulge field stars from \citet{2010A&A...513A..35A} and \citet{2014AJ....148...67J}, respectively. The error bars represent the standard deviation in the star-to-star scatter. No corrections for different adopted Solar abundance values were performed.}
    \label{fig:oddz_elements}
\end{figure}

We have found mean values for $[O/Fe]$, $[Na/Fe]$, $[Mg/Fe]$, and $[Al/Fe]$ that are compatible with other inner Galaxy GCs and field stars (Fig.~\ref{fig:oddz_elements} for odd-Z species Na and Al and Fig.~\ref{fig:alpha_elements} for $\alpha$-elements O and Mg). Most of our sample has very shallow O lines, therefore we opted to compute $[O/Fe]$ for the few with a well-defined forbidden $OI$ line at $6300.30$ \r{A}. Individual values are presented in Table~\ref{tab:abundance_means}. The sample has a star-to-star spread in these elements smaller than the uncertainties for each species (Table~\ref{tab:abundance_means2}), except for Na if star $b3$, a dramatic outlier, is considered. Thefore, within uncertainties, we do not detect a spread in these light elements save for the outlier, supporting it as a SG star. 

\begin{figure}
	\includegraphics[width=\columnwidth]{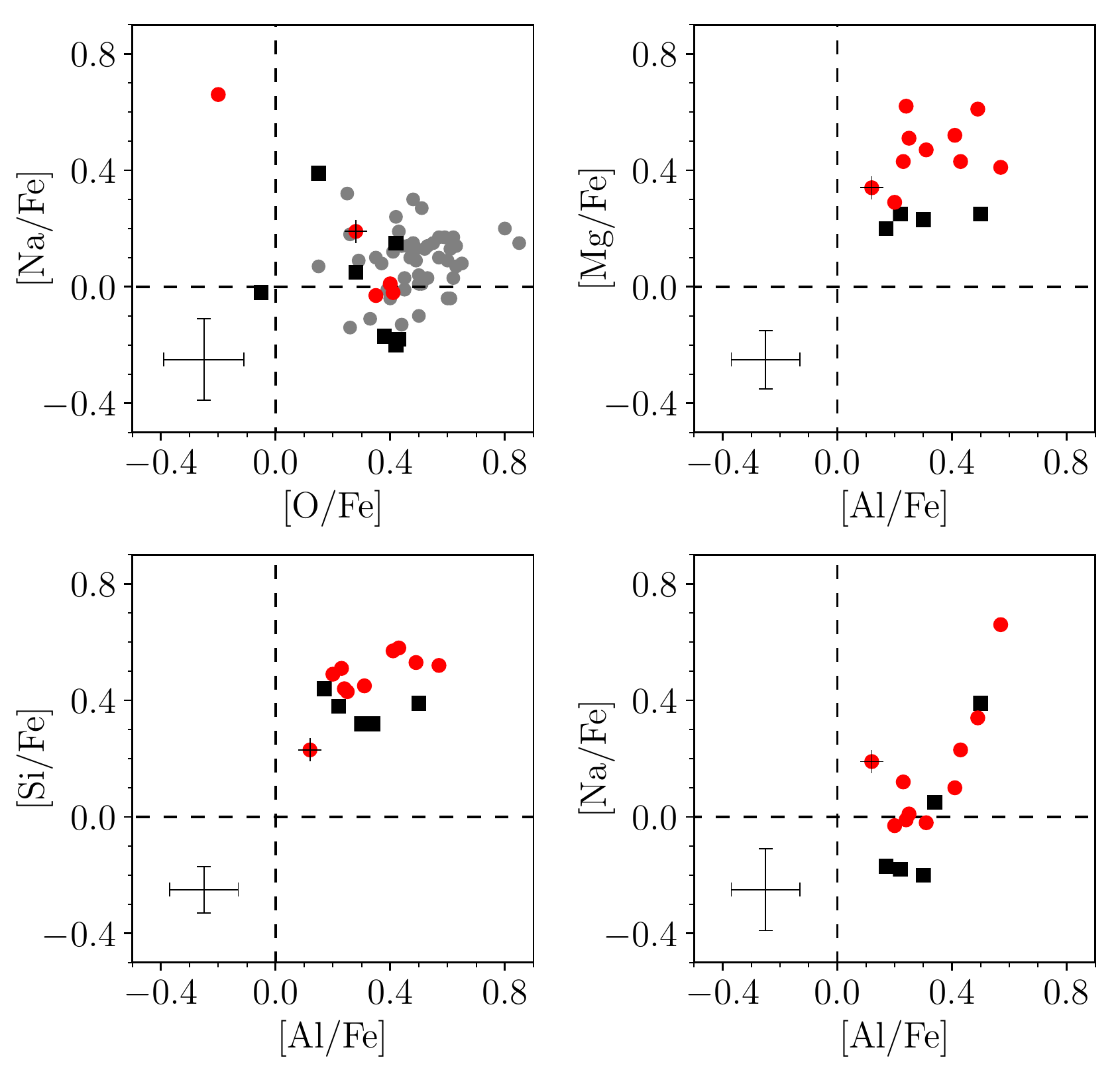}
    \caption{\emph{Red cirles:} this work, with a black cross over a red circle marking star $b6$ due to its lower SNR. \emph{Black squares:} stars 1 to 7 as computed by RA16. \emph{Grey circles:} HB stars from NGC 6723 \citep{2015A&A...573A..92G}. Typical $\sigma$ uncertainties by species for this work are presented on the lower left of each panel.}
    \label{fig:light_correlations}
\end{figure}

Furthermore, no Mg-Al anticorrelation is evident (Fig.~\ref{fig:light_correlations}). We have computed a Pearson correlation test between these two species, finding no statistically significant (anti-)correlation both with and without star $b3$. The absence of a spread in Al is consistent with expectations of AGB nucleosynthesis, where models predict that a higher metallicity correlates with a more modest Al yield, with little production in stars more metal-rich than $[\ion{Fe}/\ion{H}] \approx -0.7$ dex \citep[e.g.][]{2008MNRAS.385.2034V,2014PASA...31...30K}. Therefore, a small Al spread supports our $[\ion{Fe}/\ion{H}]$ results. 

Particularly high temperatures in the hot bottom burning of AGB stars can result in the peculiar GC anticorrelations between C and N, O and Na, and Mg and Al \citep[][and references therein]{2014PASA...31...30K}. This scenario would also carry a signature of breakout in the MgAl cycle in the form of an Al-Si correlation, with Al rich stars also present enrichment in Si. We have found no statistically significant correlation as most stars concentrate tightly around the mean, though the two stars that deviate more from the mean (stars $b3$ and $b6$) do present the expected pattern of being either enriched or depleted in both species at the same time. 

Of the stars with available $[O/Fe]$ in our sample, four are firmly within the definition of FG stars according to the criteria developed by \citet{2009A&A...505..117C}, while star $b3$ presents the high $[Na/Fe]$ and low $[O/Fe]$ typical of SG stars. Among the stars without a $[O/Fe]$ measurement, none has high enough Na to be considered a SG star. Stars $b1$ and $b4$ have been studied photometrically by \citet{2016ApJ...832...99L} and confirmed to be CN-weak, that is, they belong to an earlier generation when compared to CN-strong stars. In the same work, the earlier generation of stars seems to be more centrally concentrated, at odds with expectations from the multiple stellar population scenario.

\subsection{Fe-peak elements} \label{sect:fe_peak_elements}
Species from the so-called Fe-peak group are still poorly understood. Transitions that result in the most easily observable lines for heavy odd-Z species are subjected to strong hyperfine splitting that must be taken into careful account in the computation of abundances, lest they end up sometimes severely overestimated or suggesting spurious trends \citep{2000ApJ...537L..57P}. Their formation sites are varied and details of their production uncertain, with different isotopes being formed at various rates in different reactions \citep{2002RvMP...74.1015W}. Depending on the metallicity range, non-LTE corrections are necessary for comparisons with SNe yield models \citep[e.g.][for Mn, Co, and Cr, respectively]{2008A&A...492..823B,2010MNRAS.401.1334B,2010A&A...522A...9B}.

\begin{figure}
	\includegraphics[width=\columnwidth]{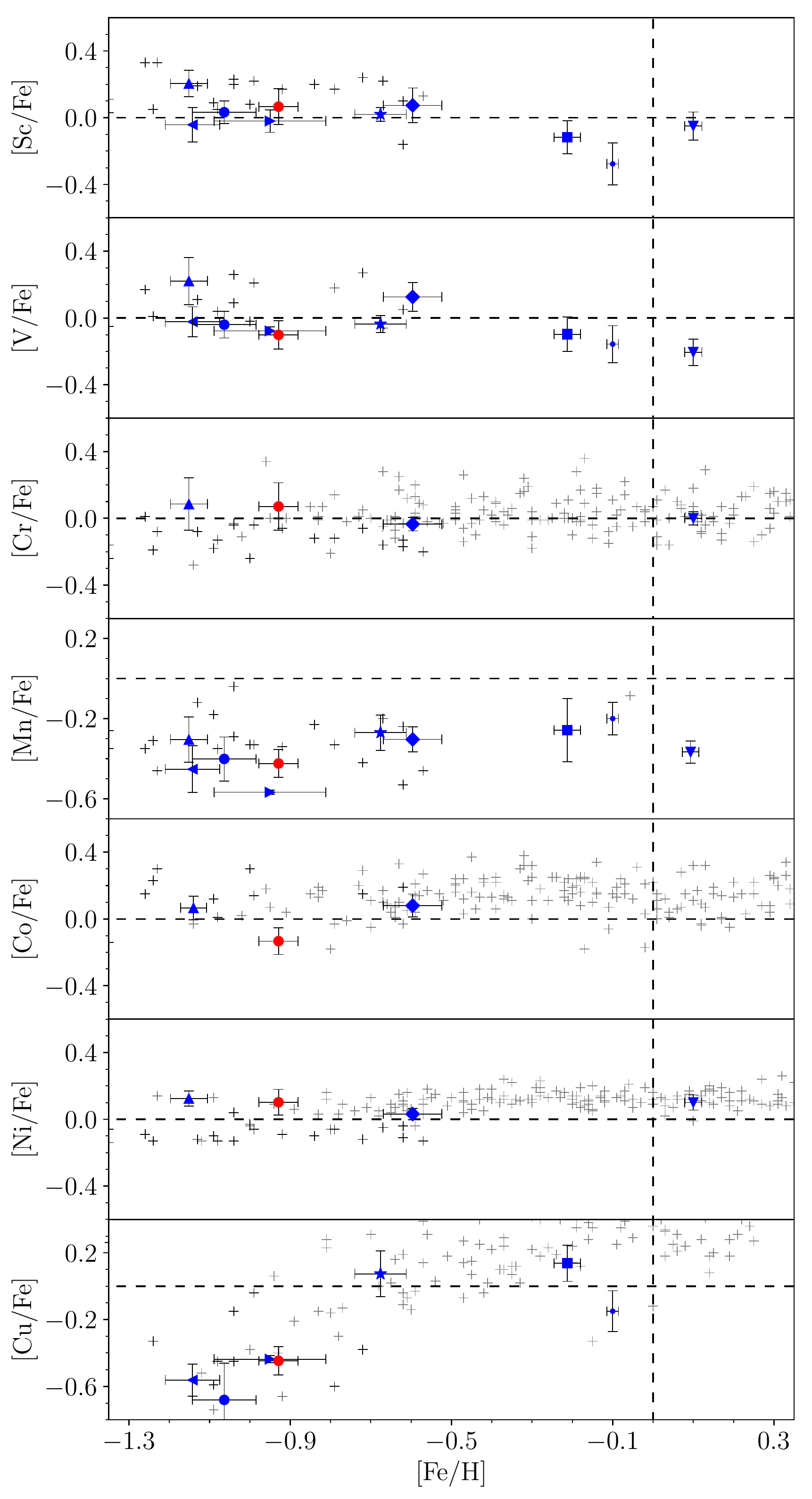}
    \caption{\emph{Red cirles:} this work. \emph{Black square}: stars 1 to 7 as computed by RA16. \emph{Blue triangle:} NGC 6266 \citep[M62,][]{2014MNRAS.439.2638Y}. \emph{Blue upside down triangle:} NGC 6528 \citep{2001AJ....122.1469C}. \emph{Blue diamond:} NGC 6366 \citep{2018MNRAS.476..690P}. \emph{Blue star, square, dot, circle, triangle to the right, triangle to the left:} 47 Tucanae, NGC 6553, NGC 6528, HP 1, NGC 6552, NGC 6558, respectively \citep{2018arXiv180106157E}. \emph{Grey crosses:} Bulge field stars from \citet{2013A&A...559A...5B} (Mn) and \citet{2014AJ....148...67J} (Cr, Co, and Ni). \emph{Black crosses:} inner Halo field stars \citep{2012ApJ...753...64I}. The error bars represent the standard deviation in the star-to-star scatter. No corrections for different adopted Solar abundance values were performed.}
    \label{fig:fepeak_elements}
\end{figure}

In the literature, GC abundance values are often reported without non-LTE corrections, and so we opted to present the results for NGC~6723 in the same manner for comparison purposes (Fig.~\ref{fig:fepeak_elements}). The coherent behavior of NGC 6723 regarding $\alpha$-elements extends to Fe-peak elements as well. While Co is slightly underabundant when compared to other GCs, it does agree with core-collapse SNe models that predict $[Co/Fe] \approx -0.1$, therefore the higher abundance trend set in the literature may be an artifact of computations that do not take HFS effects into account. The clear separation from 47 Tucanae in Cu despite the proximity in $[\ion{Fe}/\ion{H}]$ is expected by the dependency the production of this species has on metallicity \citep{2006ApJ...653.1145K,2006NuPhA.777..424N}.

\subsection{Neutron-capture elements}
As with Fe-peak species, neutron($n$)-capture elements boast strong HFS effects in their line formation and data about them are relatively sparse. They are produced in varied sites as heavy nuclei capture neutrons that, if unstable, are transformed to protons via $\beta$ decay. In neutron-poor environments where the neutron capture timescale is longer than the $\beta$ decay timescale, the slow($s$)-process dominates. While most $n$-capture elements can be produced by either process, the $s$-process dominates the production of Ba, and the $r$-process the production of Eu, at least in Solar studies. Therefore, these two species are proposed as tracers of the two nucleosynthetic mechanisms.

\begin{figure}
	\includegraphics[width=\columnwidth]{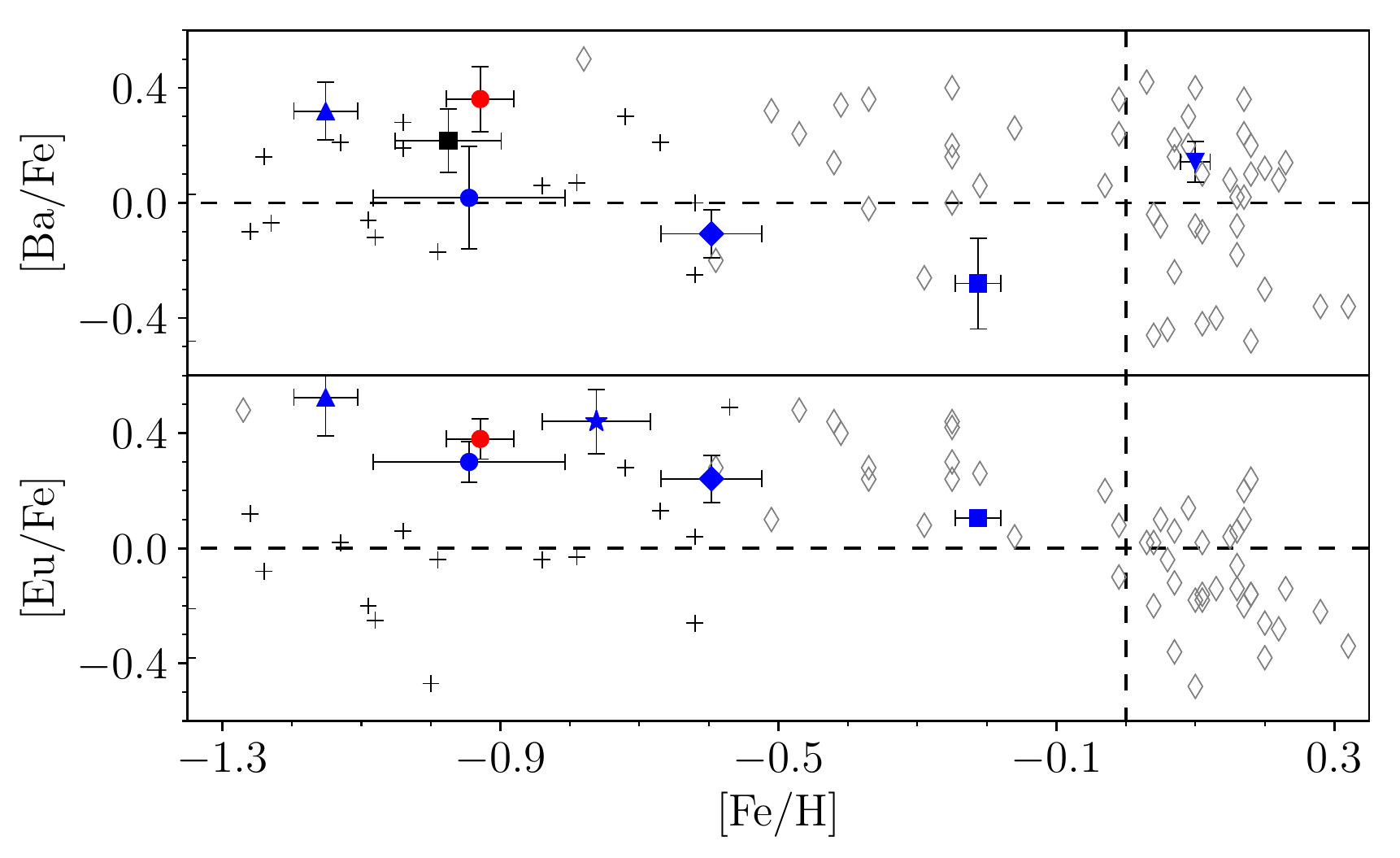}
    \includegraphics[width=\columnwidth]{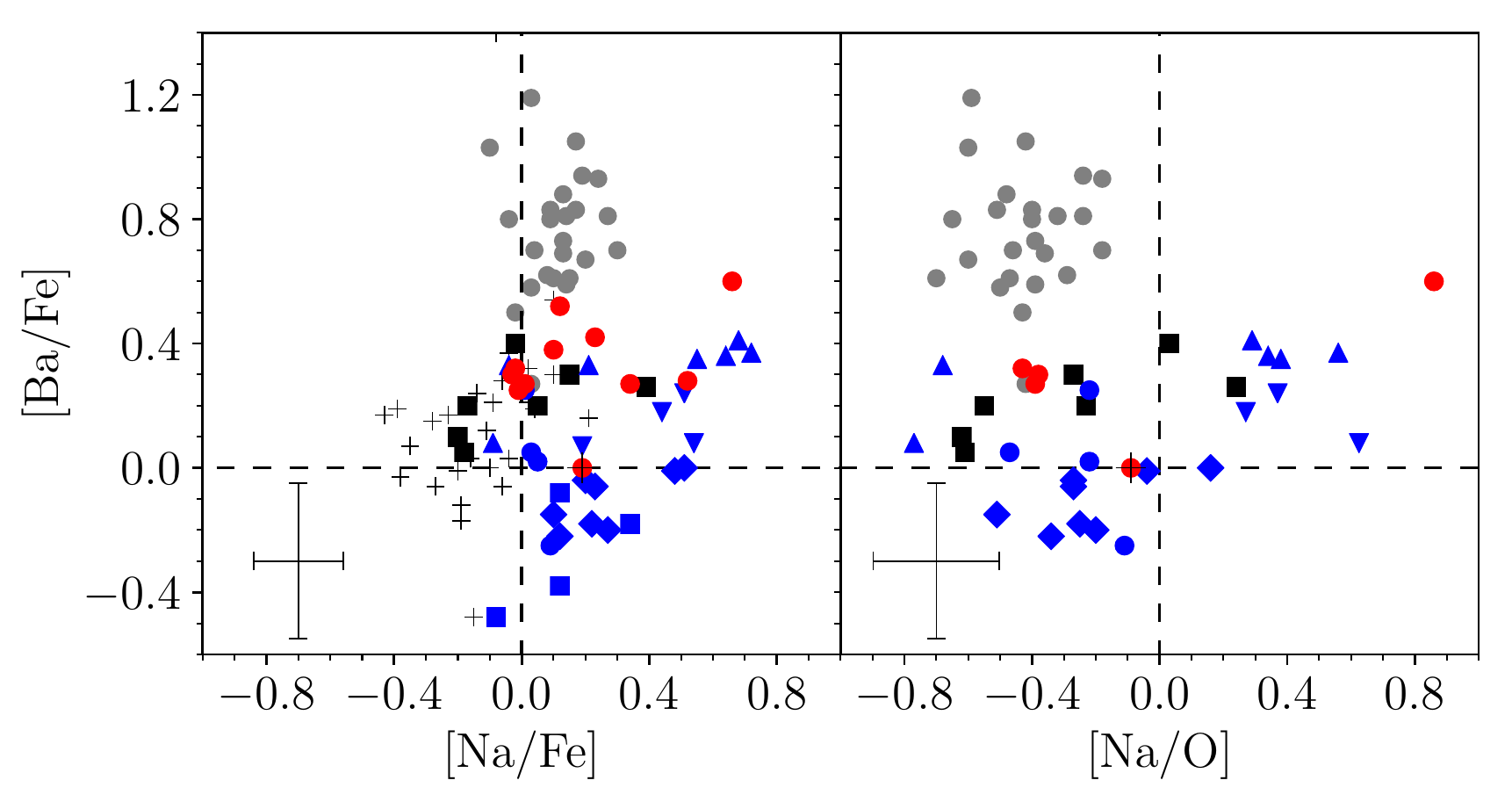}
    \caption{\emph{Red cirles:} this work, with a black cross over a red circle marking star $b6$ due to its lower SNR. \emph{Black square}: stars 1 to 7 as computed by RA16. \emph{Grey circles:} HB stars from NGC 6723 \citep{2015A&A...573A..92G}. \emph{Blue triangle:} NGC 6266 \citep[M62,][]{2014MNRAS.439.2638Y}. \emph{Blue upside down triangle:} NGC 6528 \citep{2001AJ....122.1469C}. \emph{Blue diamond:} NGC 6366 \citep{2018MNRAS.476..690P}. \emph{Blue star:} 47 Tucanae \citep{2014ApJ...780...94C}. \emph{Blue square:} NGC 6553 \citep{2006A&A...460..269A}. \emph{Blue circle:} NGC 6522 \citep{2014A&A...570A..76B}. \emph{Grey diamonds:} Bulge field giants \citep{2016A&A...586A...1V}. \emph{Black crosses:} inner Halo field stars \citep{2012ApJ...753...64I}. In the top panel, the error bars represent the standard deviation in the star-to-star scatter. In the bottom panel, errors bars represent the typical $\sigma$ uncertainties by species for this work. No corrections for different adopted Solar abundance values were performed.}
    \label{fig:ncapture_elements} 
\end{figure}

Most studies to date have employed dwarfs and/or local disk stars. As can be seen in the top panel of Fig.~\ref{fig:ncapture_elements}, results for Eu and Ba present a large scatter even when restricted to GCs and Bulge and Halo stars. Interestingly, \citet{2015A&A...573A..92G} identified a radial gradient in $[Ba/Fe]$ in their sample of $30$ red HB stars, and a remarkably high mean abundance ($[Ba/Fe] = 0.75$ dex). The authors point out that this high value may be due to systematics, which seems likely considering the value found in this work ($[Ba/Fe] = 0.36 \pm 0.11$ dex) and by \citet{2016A&A...587A..95R} ($0.22 \pm 0.21$ dex), as displayed in Fig.~\ref{fig:ncapture_elements}. This n-capture element is known to be subjected to NTLE effects for cool metal-poor stars, but the effect requires a correction of only about $-0.10$ to $-0.15$ dex for the line and atmospheric parameters range for the stars in \citet{2015A&A...573A..92G} considering the maximum value of $[Ba/Fe] = 0.4$ dex included in the grid of theoretical NTLE corrections computed by \citet{2015A&A...581A..70K}. The value of the correction is similar for the lines and ranges of atmospheric parameters covered by the stars in our sample.

As the $Ba$ lines in our spectra were strong, we applied the same method to Arcturus to test their reliability, finding a value of $-0.25$ dex with a line-by-line standard deviation of $0.07$ dex, in agreement with the results of \citet{2009MNRAS.400.1039W} ($[Ba/Fe] = -0.19 \pm 0.08$ dex). Unfortunately, due to our small sample size that contains stars mostly about the same radial distance from the GC centre, we are unable to investigate the presence of a radial gradient.

\section{Overview} \label{sect:overview}

\subsection{Inner Halo membership}
Cross-matching the coordinates of our sample with the Gaia DR2 catalog, we collected the proper motions of all stars. Their cartesian spatial velocities are all within a tight range, with $(U,V,W) = (-114.6, 155.5, -33.7)$ kms$^{-1}$, and a star-to-star standard deviation $(\sigma_U, \sigma_V, \sigma_W) = (4.7, 4.4, 4.0)$ kms$^{-1}$. These results are in excellent agreement with the estimate of \citet{2018arXiv180709775V} for NGC 6723 using Gaia DR2 data, which results in $(U,V,W) = (-101.0, 155.6, -35.1)$ kms$^{-1}$. We derived these velocities considering a Solar pecular motion $(U,V,W)_{\odot} = (11.1, 12.2, 7.3)$ kms$^{-1}$ \citep{2010MNRAS.403.1829S}, and a local standard of rest with $V_{LSR} = 220$ kms$^{-1}$ at a Galactocentric distance $R_{GC} = 8.0$ kpc \citep{2003AJ....125.1373D}.

At about $2.6$ kpc from the Galactic centre \citep[][2010 edition]{1996AJ....112.1487H}, NGC 6723 lies at the edge of the inner Halo and the outer Bulge. \citet{2003AJ....125.1373D} established that it has a nearly polar orbit, a result confirmed by \citet{2018arXiv180709775V} with Gaia DR2 data, reinforcing its inner Halo membership. Using state-of-the-art proper motions and radial velocities from a variety of catalogs spanning up to $65$ years, \citet{2018AstBu..73..162C} computed orbits for both an axisymmetric and a barred Galactic potential for $115$ GCs. Their results reproduce an eccentric and sharply polar orbit for NGC 6723. For a barred potential and considering $R_{GC} = 8.3$ kpc, they found a mean and a maximum apocentre at $2.3$ and $3.7$ kpc, respectively, while the mean and minimum pericentre are at $0.5$ and $0.0$ kpc, respectively. 

Therefore, NGC 6723 is known to dive deeply into the Bulge, but it is able to reach distances that can be considered to be beyond it depending on how one postulates its spatial dimensions \citep[see][for a recent review of the Bulge]{2018arXiv180501142B}. Given its polar orbit, it is unlikely to belong to a rotationally supported system such a pseudo-Bulge, although a classic Bulge, being pressure supported, could contain an object with such kinematics. Whether NGC 6723 is native to the inner Bulge and was ejected to slightly greater distances, or the reverse, is unclear, but the updated proper motions of Gaia DR 2 confirm that it is a permanent resident of the inner region of the Milky Way.

\subsection{Horizontal branch morphology}

The color-magnitude diagram of NGC 6723 displays an extended HB, typical of intermediate to metal-poor GCs. It has long been known that GCs with similar metallicity can, however, have quite different HB morphologies, a fact known as the second parameter phenomenon. This second parameter involves a combination of many different characteristics, such as total GC luminosity (and therefore mass), age, and helium abundance variations \citep[e.g.][and references therein]{1994ApJ...423..248L,2006A&A...452..875R,2009Ap&SS.320..261C}. 

\citet{2015A&A...573A..92G} found 43 stars in a sample of 47 HB stars in a wide range of color presented chemical abundances compatible with a first stellar generation, and only the warmest four stars could belong to a second generation. While this result reinforces that stars of different generations populate different regions of the HB, the spread in color for the chemically homogeneous first generation is yet to be understood.

\citet{2014ApJ...785...21M}, using HST photometry, adopted two variables to describe the morphology of the HB: the distance in color between the redder end of the HB and the RGB (L1), and the length in color of the HB (L2). Considering their sample and parameters, NGC 6723 is morphologically similar to NGC 6362 ($[\ion{Fe}/\ion{H}] = -0.99$), NGC 7006 ($[\ion{Fe}/\ion{H}] = -1.52$), and NGC 6715 ($[\ion{Fe}/\ion{H}] = -1.49$). On the other hand, NGC 6652 and Palomar 12 have predominantly red HBs despite metallicity values close to that of NGC 6723 ($[\ion{Fe}/\ion{H}] = -0.85$ and $-0.81$, respectively). 

\begin{figure}
    \includegraphics[width=\columnwidth]{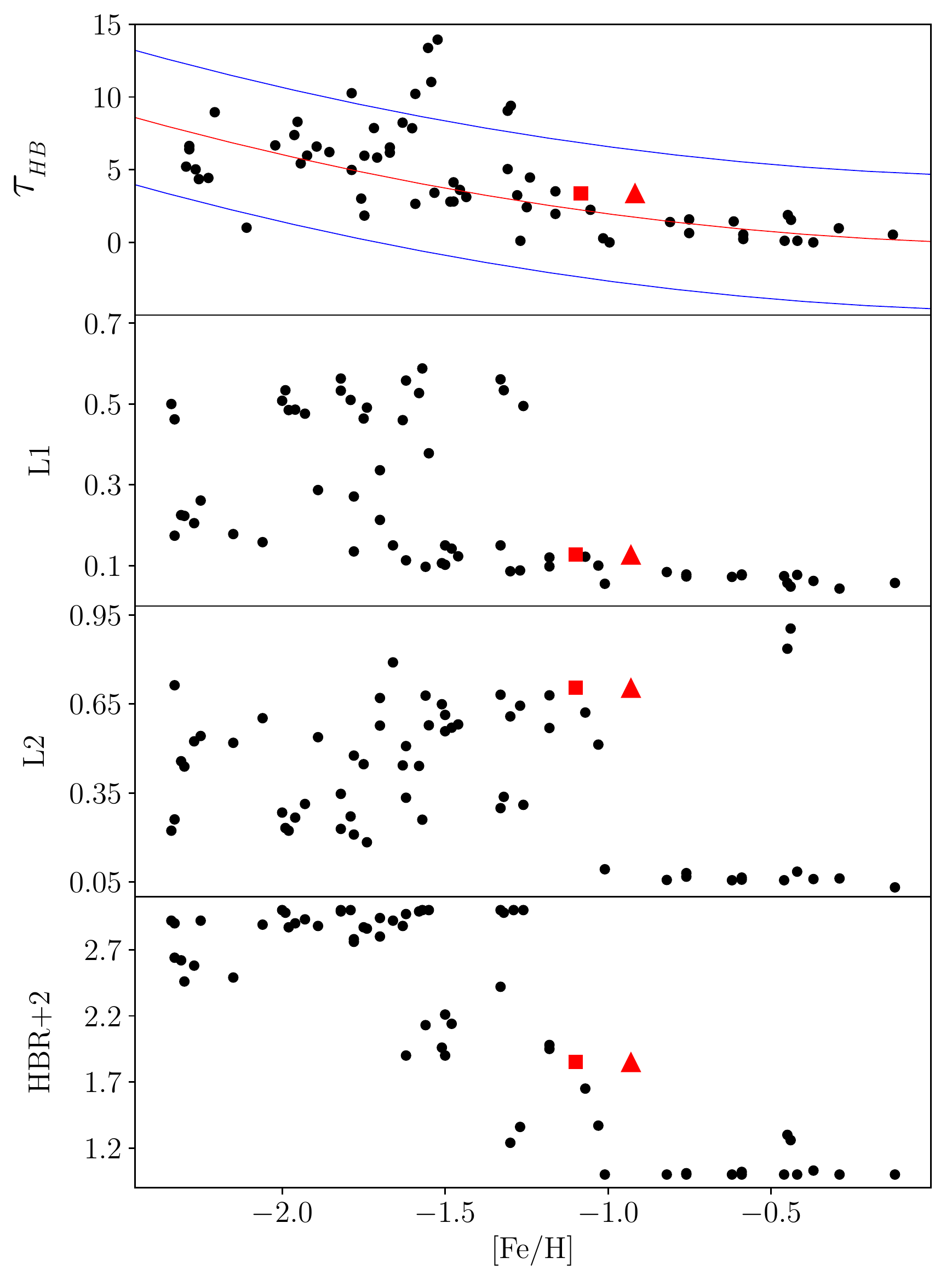}
    \caption{Different HB morphology parameters as a function of metallicity: $\tau_{HB}$ index \citep{torellisubmitted} with a red line indicating the best fit and blue lines the $1.5\sigma$ levels, L1 and L2 \citep{2014ApJ...785...21M}, and the HBR \citep{1988csa..proc..149L}. \emph{Black cirles:} GCs with available $\tau_{HB}$ index, with metallicities from \citep{2009A&A...508..695C}. \emph{Red square:} NGC~6723 considering $[Fe/H] = -1.10$ dex. \emph{Red triangle:} NGC~6723 considering the metallicity found in this work.}
    \label{fig:HB_morphology} 
\end{figure}

It is evident that L1 is strongly influenced by metallicity, with two well-defined groups for $[\ion{Fe}/\ion{H}] < -1$ and exclusively small values of L1 for their sample of more metal-rich GCs. Note that Lyng\aa 7, a metal-rich GC, appears mixed with the low-metallicity group. However, both \citet{2016A&A...590A...9D} and \citet{2004AJ....128.1228S} agree on a higher metallicity, $[\ion{Fe}/\ion{H}] = -0.61 \pm 0.10$ and $-0.76 \pm 0.06$, respectively, putting it firmly alongside the other metal-rich GCs. Our metallicity value for NGC 6723 puts it right at the edge between the two small L1 groups, creating a vast range of metallicities for very similar values of L1 (Fig.~\ref{fig:HB_morphology}). For L2, the sharp contrast between the metal-rich and metal-poor groups remains unaffected, suggesting that at $[\ion{Fe}/\ion{H}] \approx -1$, there is a regime change from the presence of a red HB at high metallicities, to a wide variety of morphologies with both red and extended HBs in lower metallicities, with NGC 6723 right at the edge of this  change. 

Like the color-based L1 and L2 parameters, star-count based HB morphology indexes such as the HBR \citep{1988csa..proc..149L} are also affected by a degeneracy in metallicity \citep[e.g.][and references therein]{1993ASPC...48..131B}, presenting very similar values for GCs with significantly different HBs. \cite{torellisubmitted} introduced the new index $\tau_{HB}$, based on the Cumulative Distribution Functions in the $(V-I)$ color and $I$-band magnitude of HB stars, which shows a quadratic correlation with $[Fe/H]$ and presents a high degree of sensitivity to morphology differences. They applied this new index to a sample of 64 GCs with homogeneous and high quality ground and ACS/HST photometry which included NGC 6723. With our metallicity value, NGC 6723 stays within $0.55 \sigma$ of the quadratic correlation between $\tau_{HB}$ and $[Fe/H]$, therefore presenting a very typical metal intermediate HB morphology. In this new index, its closest neighbor in both HB morphology and metallicity is still the less massive NGC 6362.

\subsection{Multiple stellar populations}
As discussed in Sect.~\ref{sect:oddz_elements}, the presence of chemical inhomogenities in GCs appears ubiquitous. NGC 6723 is no exception. Using a "Ca+CN" photometric filter, \citet{2016ApJ...832...99L} observed $596$ RGB stars, finding a SG/FG ratio of $0.62$. \citet{2017MNRAS.464.3636M} found a similar ratio of $0.64$ in $695$ RGB stars employing a method of pseudo-two-color diagram with \emph{Hubble Space Telescope} UV photometry, and also verified that, unlike some GCs in their sample, NGC 6723 has two well defined and internally homogeneous populations. This SG/FG ratio is typical: the mean value found by \citet{2017MNRAS.464.3636M} lies at $0.62 \pm 0.13$ in a sample of $57$ GCs. With a sample of $33$ GCs, \citet{2015MNRAS.453..357B} reached a similar result, with the mean at $0.68 \pm 0.07$.

Spectroscopically, \citet{2015A&A...573A..92G} studied a sample of $43$ HB stars and found that only $4$ (SG/FG $= 0.09$) of them had a possible SG membership. This value is a lower limit due to the segregation expected between FG and SG stars along the HB, which means SG stars are more likely to be found at temperatures higher than those covered by their observations. In our sample, only $1$ star is likely to belong to the SG. \citet{2015MNRAS.453..357B} present the argument that photometric and spectroscopic criteria to separate different populations may not be in agreement. While this possibility must be studied further with large homogenized samples with both spectroscopic and photometric measurements, this apparent disagreement in NGC 6723 is likely due to small number statistics and, in the case of HB stars, a selection bias.

In this work, star $b3$ is the only one with a pattern of Na enrichment and O depletion. Two stars in our sample, $b1$ and $b4$, appear as Na-depleted and O-enhanced in our study, and are recognized as CN-weak photometrically by \citet{2016ApJ...832...99L}, confirming them as belonging to the so-called FG.

\subsection{Abundance pattern}

\begin{figure}
	\includegraphics[width=\columnwidth]{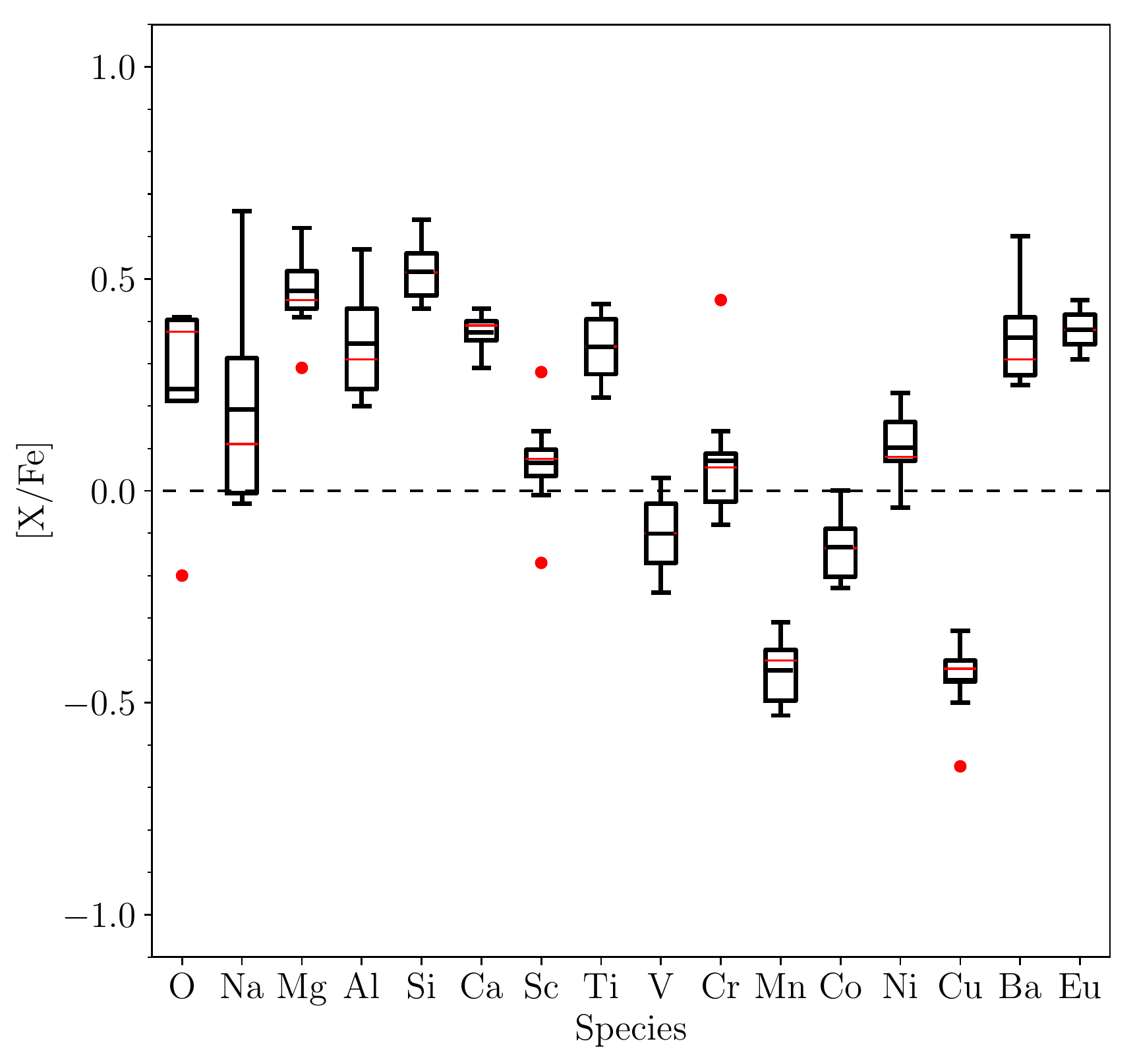}
    \caption{Final abundance pattern for NGC 6723. Boxes represent the interquartile ranges (IQR), whiskers the $1.5 \times$IQR limits (approximately $2.7 \times$ the standard deviation in the star-to-star spread $\sigma_*$), black lines the means, red lines the medians, red circles the outliers.}
    \label{fig:means_boxplot}
\end{figure}

Galactic GCs are known to present sharply different behaviors in the $\alpha$-element versus $[Fe/H]$ plane, similar to the "knee" found in field stars although at different values. Our results place NGC 6723 alongside the metal-poor GCs in this plane. Overall, this metal-intermediate GC lies at the edge between the two metallicity regimes and is more similar to its lower metallicity counterparts, without any chemical peculiarities. An overview of means and star-to-star spread for all species is shown in Fig.\ref{fig:means_boxplot} and Table~\ref{tab:abundance_means2}, while Table~\ref{tab:abundance_means} displays abundances for individual stars.

\begin{table*}
	\centering
	\caption{Final abundances for the analyzed stars.}
	\label{tab:abundance_means}
	\begin{tabular}{cccccccccccc} 
		\hline
		$[X/Fe]$ & $a1$ & $a2$ & $a3$ & $a4$ & $b1$ & $b2$ & $b3$ & $b4$ & $b5$ & $b6$ & $b7$ \\ 
		\hline
        \ion{O}{I}    & --       & --      & --      & --      & $+0.41$ & $+0.40$ & $-0.20$ & $+0.35$ & --      & $+0.28$ & --       \\ 
        \ion{Na}{I}   & $+0.12$  & $+0.10$ & $+0.23$ & $+0.52$ & $-0.02$ & $+0.01$ & $+0.66$ & $-0.03$ & $+0.34$ & $+0.19$ & $-0.01$  \\ 
        \ion{Mg}{I}   & $+0.43$  & $+0.52$ & $+0.43$ & $+0.43$ & $+0.47$ & $+0.51$ & $+0.41$ & $+0.29$ & $+0.61$ & $+0.34$ & $+0.62$  \\ 
        \ion{Al}{I}   & $+0.23$  & $+0.41$ & $+0.43$ & --      & $+0.31$ & $+0.25$ & $+0.57$ & $+0.20$ & $+0.49$ & $+0.12$ & $+0.24$  \\ 
        \ion{Si}{I}   & $+0.52$  & $+0.57$ & $+0.58$ & $+0.64$ & $+0.45$ & $+0.43$ & $+0.52$ & $+0.49$ & $+0.53$ & $+0.23$ & $+0.44$  \\ 
        \ion{Ca}{I}   & $+0.40$  & $+0.43$ & $+0.41$ & $+0.39$ & $+0.29$ & $+0.37$ & $+0.40$ & $+0.35$ & $+0.39$ & $+0.14$ & $+0.30$  \\ 
        \ion{Sc}{II}  & $+0.09$  & $+0.10$ & $+0.14$ & $+0.28$ & $+0.06 $& $-0.01$ & $+0.05$ & $+0.09 $& $+0.03$ & $-0.08$ & $-0.17$  \\ 
        \ion{Ti}{I}   & $+0.30$  & $+0.41$ & $+0.39$ & $+0.44$ & $+0.27$ & $+0.38$ & $+0.27$ & $+0.29$ & $+0.43$ & $+0.34$ & $+0.22$  \\ 
        \ion{V}{I}    & $-0.10$  & $-0.15$ & $-0.03$ & --      & $-0.06$ & $-0.01$ & $-0.18$ & $-0.17$ & $+0.03$ & --      & $-0.24$  \\ 
        \ion{Cr}{I}   & $+0.04$  & $+0.09$ & $-0.04$ & $+0.08$ & $+0.14$ & $+0.07$ & $+0.45$ & $-0.01$ & $-0.03$ & --      & $-0.08$  \\ 
        \ion{Mn}{I}   & $-0.50$  & $-0.40$ & $-0.48$ & $-0.50$ & $-0.39$ & $-0.37$ & $-0.40$ & $-0.36$ & $-0.31$ & $-0.35$ & $-0.53$  \\ 
        \ion{Co}{I}   & $-0.22$  & --      & $+0.00$ & --      & $-0.15$ & $-0.12$ & $-0.23$ & --      & $-0.08$ & $-0.18$ & --       \\ 
        \ion{Ni}{I}   & $+0.08$  & $+0.20$ & $+0.19$ & $+0.23$ & $+0.07$ & $+0.06$ & $+0.07$ & $+0.08$ & $+0.08$ & $+0.11$ & $-0.04$  \\
        \ion{Cu}{I}   & $-0.45$  & $-0.40$ & $-0.33$ & --      & $-0.42$ & $-0.42$ & $-0.50$ & $-0.45$ & $-0.40$ & --      & $-0.65$  \\  
        \ion{Ba}{II}  & $+0.47$  & $+0.40$ & $+0.40$ & $+0.20$ & $+0.32$ & $+0.27$ & $+0.60$ & $+0.30$ & $+0.25$ & --      & $+0.25$  \\ 
        \ion{Eu}{II}  & $+0.45$  & --      & --      & --      & --      & --      & --      & $+0.31$ & --      & --      & --       \\ 
		\hline
	\end{tabular}
\end{table*}

\begin{table}
	\centering
	\caption{Mean abundances for the analyzed stars excluding $b6$. For the light elements O, Na and Al, the two values for mean and $\sigma_*$ represent those values computed with (first line) and without (second line) star $b3$.}
	\label{tab:abundance_means2}
	\begin{tabular}{cccr}
		\hline
		$[X/Fe]$ & Mean & $\sigma_*$ & $\sigma$ \\
		\hline
        \ion{O}{I}    & $+0.24$  &   $0.26$      & $0.14$ \\
                      & $+0.39$  &   $0.03$      & $0.14$ \\
        \ion{Na}{I}   & $+0.19$  &   $0.23$      & $0.14$ \\
                      & $+0.14$  &   $0.18$      & $0.14$ \\
        \ion{Mg}{I}   & $+0.47$  &   $0.09$      & $0.10$ \\
        \ion{Al}{I}   & $+0.35$  &   $0.12$      & $0.12$ \\
                      & $+0.32$  &   $0.10$      & $0.12$ \\
        \ion{Si}{I}   & $+0.52$  &   $0.06$ 	 & $0.08$ \\
        \ion{Ca}{I}   & $+0.37$  &   $0.04$ 	 & $0.20$ \\
        \ion{Sc}{II}  & $+0.07$  &   $0.11$ 	 & $0.15$ \\
        \ion{Ti}{I}   & $+0.34$  &   $0.07$ 	 & $0.24$ \\
        \ion{V}{I}    & $-0.10$	 &   $0.08$	 & $0.30$ \\
        \ion{Cr}{I}   & $+0.07$	 &   $0.14$ 	 & $0.15$ \\
        \ion{Mn}{I}   & $-0.42$	 &   $0.07$ 	 & $0.21$ \\
        \ion{Co}{I}   & $-0.13$	 &   $0.08$	 & $0.18$ \\
        \ion{Ni}{I}   & $+0.10$	 &   $0.08$ 	 & $0.13$ \\
        \ion{Cu}{I}   & $-0.45$	 &   $0.08$ 	 & $0.18$ \\
        \ion{Ba}{II}  & $+0.36$	 &   $0.11$ 	 & $0.25$ \\
        \ion{Eu}{II}  & $+0.38$	 &   $0.07$ 	 & $0.15$ \\
		\hline
	\end{tabular}
\end{table}

\section{Conclusions} \label{sect:conclusions} 
We have performed a high resolution spectral analysis of eleven RGB stars in the old GC NGC 6723. It is located at a region of low interstellar reddening and has an inner Halo orbit that crosses the Bulge, with an apocenter of less than $4$ kpc. We found an intermediate metallicity $[\ion{Fe}/\ion{H}] = -0.93$, with a star-to-star spread of $\sigma_* = 0.05$ dex. These values were computed excluding star $b6$ due to its lower SNR and likely spurious metallicity value of $[\ion{Fe}/\ion{H}] = -0.70$ dex.

Several chemical species were employed in order to expand our understanding of this sparsely studied object, with a special attention regarding Fe-peak and $n$-capture elements as those have few measurements in GCs in general. Overall, the results for NGC 6723 are in agreement with other GCs of similar metallicity for all species. In the $\alpha$-element versus $[Fe/H]$ plane, where there is a sharp difference between metal-poor and metal-rich GCs, NGC 6723 follows the trend set by the former. The extended HB and the presence of numerous RR Lyrae stars also make this metal-intermediate GC a closer relative to its lower metallicity counterparts.

Considering the multiple stellar generations scenario, in our sample only one star ($b3$) displays typical second generation chemistry, with extreme values of both Na and O. Removing this one star, we find no spread in light odd-Z elements Al and Na greater than uncertainties. Two stars in our sample ($b1$ and $b4$) have been studied by \citet{2016ApJ...832...99L} with CN-sensitive photometric filters, and we confirm their result that they belong to a primordial generation. We could not confirm the high $Ba$ abundance found by \citet{2015A&A...573A..92G} ($[Ba/Fe] = 0.75$ dex), with the values of this work pointing to a mean of $0.36 \pm 0.11$ dex for this n-capture element, in agreement with the estimate of \citet{2016A&A...587A..95R}.

With a significantly extended HB, NGC 6723 can be mistaken for a metal-poor GC based on photometry alone. However, we confirm the higher metallicity found by several previous works, although only one of them using high resolution spectroscopy. The intermediate metallicity we derived puts NGC 6723 right at the edge between the low and high metallicy regimes that correlate strongly with HB morphology (Fig.~\ref{fig:HB_morphology}). Considering these results and the work of \citet{2015A&A...573A..92G} that found a large color spread among chemically homogeneous HB stars in this GC, NGC 6723 is an important laboratory to explore the so-called second parameter problem regarding the varied HB morphologies in GCs.

\section*{Acknowledgements}

This research has been developed with financial support from CNPq (Brazil). It has employed the NASA Astrophysics Data System Bibliographic Services. We thank the referee, Dr. R. G. Gratton, for the detailed reading of an early version of this manuscript and for his constructive suggestions that improved its content and readability.

%%%%%%%%%%%%%%%%%%%%%%%%%%%%%%%%%%%%%%%%%%%%%%%%%%

\bibliographystyle{mnras}
\bibliography{ngc6723bibliography}

%%%%%%%%%%%%%%%%%%%%%%%%%%%%%%%%%%%%%%%%%%%%%%%%%%

%%%%%%%%%%%%%%%%% APPENDICES %%%%%%%%%%%%%%%%%%%%%

%\appendix


%%%%%%%%%%%%%%%%%%%%%%%%%%%%%%%%%%%%%%%%%%%%%%%%%%

% do not change these lines
\bsp	% typesetting comment
\label{lastpage}
\end{document}